\documentclass[twocolumn]{aastex61}

\usepackage{graphicx}
\usepackage{epstopdf}
\usepackage{verbatim}
\usepackage{color}
\usepackage{amsmath}

\usepackage[]{hyperref}

\hypersetup{%
  colorlinks=true,
  linkcolor=red,
  citecolor=blue,
  bookmarksopen=true,
  bookmarksnumbered=true,
  pdfstartpage={1},  
  pdftitle = {},
  pdfsubject = {},
  pdfauthor = {} ,
  pdfkeywords = {},
  pdfproducer = {LaTeX with hyperref}
}

\def\Bmax{\ifmmode{\>\vert B\vert_{\text{max}}}\else{$\vert B\vert_{\text{max}}$}\fi}
\def\Bnine{\ifmmode{\>\vert B\vert_{90}}\else{$\vert B\vert_{90}$}\fi}

\def\nH{\ifmmode{\>n_{\textnormal{\sc h}}} \else{$n_{\textnormal{\sc h}}$}\fi}

\def\mG{\ifmmode{\>\mu\mathrm{G}}\else{$\mu$G}\fi}
\def\erg{\ifmmode{\> {\rm erg}}\else{erg}\fi}
\def\keV{\ifmmode{\> {\rm keV}}\else{keV}\fi}

\def\deg{\ifmmode{\>^{\circ}}\else{$^{\circ}$}\fi}
\def\onedeg{\ifmmode{\>1^{\circ}}\else{$1^{\circ}$}\fi}

\def\xvir{\ifmmode{\>x_{vir}}\else{$x_{vir}$}\fi}
\def\Mvir{\ifmmode{\>M_{vir}}\else{$M_{vir} $}\fi}
\def\rvir{\ifmmode{\>r_{vir}}\else{$r_{vir}$}\fi}
\def\vvir{\ifmmode{\>v_{vir}}\else{$v_{vir}$}\fi}

\def\tratio{\ifmmode{\>\tau}\else{$\tau$}\fi}

\def\rms{\ifmmode{\>r_{\textnormal{\sc ms}}}\else{$r_{\textnormal{\sc ms}}$}\fi}

\def\Mpc{\ifmmode{\>{\rm Mpc}} \else{Mpc}\fi}
\def\kpc{\ifmmode{\>{\rm kpc}} \else{kpc}\fi}
\def\pc{\ifmmode{\>{\rm pc}} \else{pc}\fi}

\def\Gyr{\ifmmode{\>{\rm Gyr}} \else{Gyr}\fi}
\def\Myr{\ifmmode{\>{\rm Myr}} \else{Myr}\fi}
\def\yr{\ifmmode{\>{\rm yr}} \else{yr}\fi}
\def\pyr{\ifmmode{\>{\rm yr}^{-1}}\else{yr $^{-1}$} \fi}
\def\s{\ifmmode{\>{\rm s}}\else{s}\fi}
\def\ps{\ifmmode{\>{\rm s}^{-1}}\else{s$^{-1}$}\fi}
\def\Hz{\ifmmode{\>{\rm Hz}}\else{Hz}\fi}

\def\kms{\ifmmode{\>{\rm km\,s}^{-1}}\else{km~s$^{-1}$}\fi}

\def\K{\ifmmode{\>{\rm K}}\else{K}\fi}

\def\sr{\ifmmode{\>{\rm sr}}\else{sr}\fi}
\def\psr{\ifmmode{\>{\rm sr}^{-1}}\else{sr$^{-1}$}\fi}
\def\arcs{\ifmmode{\>{\rm arcsec}}\else{arcsec}\fi}
\def\parcs{\ifmmode{\>{\rm arcsec}^{-1}}\else{arcsec${-1}$}\fi}
\def\parcss{\ifmmode{\>{\rm arcsec}^{-2}}\else{arcsec${-2}$}\fi}

\def\cm{\ifmmode{\>{\rm cm}}\else{cm}\fi}
\def\cc{\ifmmode{\>{\rm cm}^{3}}\else{cm$^{3}$}\fi}
\def\sqc{\ifmmode{\>{\rm cm}^{2}}\else{cm$^{2}$}\fi}
\def\pcc{\ifmmode{\>{\rm cm}^{-3}}\else{cm$^{-3}$}\fi}
\def\psc{\ifmmode{\>{\rm cm}^{-2}}\else{cm$^{-2}$}\fi}

\def\g{\ifmmode{\>{\rm g}}\else{g}\fi}
\def\Msun{\ifmmode{\>{\rm M}_{\odot}}\else{M$_{\odot}$}\fi}
\def\hMsun{\ifmmode{\> h^{-1}{\rm M}_{\odot}}\else{$h^{-1}$M$_{\odot}$}\fi}

\def\Zsun{\ifmmode{\>{\rm Z}_{\odot}}\else{Z$_{\odot}$}\fi}

\def\rayl{\ifmmode{\>{\rm R}}\else{R}\fi}
\def\mR{\ifmmode{\>{\rm mR}}\else{mR}\fi}

\renewcommand{\ion}[2]{\hbox{#1\,{\sc #2}}}

\def\lya{\ifmmode{\>{\rm Ly}\alpha}\else{Ly$\alpha$}\fi}

\def\Ha{\ifmmode{\>{\rm H}\alpha}\else{H$\alpha$}\fi}
\def\Hb{\ifmmode{\>{\rm H}\beta}\else{H$\beta$}\fi}

\def\HI{\ifmmode{\> \textnormal{\ion{H}{i}}} \else{\ion{H}{i}}\fi}
\def\HII{\ifmmode{\> \textnormal{\ion{H}{ii}}} \else{\ion{H}{ii}}\fi}
\def\CIV{\ifmmode{\> \textnormal{\ion{C}{iv}}} \else{\ion{C}{iv}}\fi}
\def\SiIV{\ifmmode{\> \textnormal{\ion{S}{iv}}} \else{\ion{Si}{iv}}\fi}

\def\NHI{\ifmmode{\> {\rm N}_{\HI}} \else{N$_{\HI}$}\fi}
\def\MHI{\ifmmode{\> {\rm M}_{ \HI}} \else{M$_{\HI}$}\fi}

\def\mua{\ifmmode{\>\mu_{ \textnormal{\Ha}}}\else{$\mu_{ \textnormal{\Ha}}$}\fi}
\def\alphabha{\ifmmode{\>\alpha_{B}^{(\textnormal{\Ha})}}\else{$\alpha_{B}^{(\textnormal{\Ha})}$}\fi}

\shorttitle{Magnetized HVCs in the halo -- a new distance constraint}
\shortauthors{Gr\o nnow et al.}

\begin{document}
\title{Magnetized high velocity clouds in the Galactic halo -- a new distance constraint}

\author{Asger Gr\o nnow}
\affiliation{Sydney Institute for Astronomy, School of Physics A28, The University of Sydney, NSW 2006, Australia}

\author{Thor Tepper-Garc\'{\i}a}
\affiliation{Sydney Institute for Astronomy, School of Physics A28, The University of Sydney, NSW 2006, Australia}

\author{Joss Bland-Hawthorn}
\affiliation{Sydney Institute for Astronomy, School of Physics A28, The University of Sydney, NSW 2006, Australia}

\author{N.\ M.\ McClure-Griffiths}
\affiliation{Research School of Astronomy and Astrophysics, Australian National University, Canberra, ACT 2611, Australia}

\correspondingauthor{Asger Gr\o nnow}
\email{asger.gronnow@sydney.edu.au}

\begin{abstract}

High velocity gas that does not conform to Galactic rotation is observed throughout the Galaxy's halo. One component of this gas, \HI\  high velocity clouds (HVCs), have attracted attention since their discovery in the 1960s and remain controversial in terms of their origins, largely due to the lack of reliable distance estimates. The recent discovery of enhanced magnetic fields towards HVCs has encouraged us to explore their connection to cloud evolution, kinematics, and survival as they fall through the magnetized Galactic halo. For a reasonable model for the halo magnetic field, most infalling clouds see transverse rather than radial field lines. We find that significant compression (and thereby amplification) of the ambient magnetic field occurs in front of the cloud and in the tail of material stripped from the cloud. The compressed transverse field attenuates hydrodynamical instabilities. This delays cloud destruction, though not indefinitely. The observed $\vec{B}$ field compression is related to the cloud's distance from the Galactic plane. As a result, the observing a rotation measure signal with radio continuum polarization provides useful distance information on a cloud's location.

\end{abstract}



\section{Introduction} \label{sec:intro}
Surrounding the Milky Way, there is a population of atomic hydrogen (\HI) clouds whose velocities deviate significantly from the allowed rotational velocities of the Galaxy.  These so-called `high velocity clouds' (HVCs) were first discovered around the Milky Way by \cite{muller63} and have over the decades been revealed to be present throughout the sky \citep[e.g.][]{putman12}.  HVCs likely originate from multiple sources.  Some may have been stripped from the satellite galaxies, specifically the Magellanic Stream, as they are accreted into the halo.  Other HVCs may be the biproducts of feedback within the disk \citep[e.g.][]{lockman02,ford10,fraternali15}.  A major challenge in understanding the origin and nature of HVCs has always been the lack of distances ($d$) to these objects. This uncertainty led to the suggestion that some HVCs may be very distant ($\sim 1$ Mpc), i.e. primordial remnants of galaxy assembly \citep{blitz99}.   However, the primordial \HI\ model was undermined by the discovery of weak H$\alpha$ recombination emission from most HVCs \citep{weiner02,putman+blandhawthorn03} arising from the disk's ionizing radiation field \citep[][]{bland99,bland02}. While this is a coarse distance constraint, the detections place most of the clouds within the distance of the Magellanic Stream ($d < 100 \kpc$ over the South Galactic Pole). HVC analogs have been found around other galaxies on similar distance scales \citep[e.g.][]{thilker04,putman09,lehner12}. Tight constraints on the distances to a handful of HVC complexes have been found through a powerful technique based on the lack of absorption features along the sightline to stars in the Milky Way halo \citep[e.g.,][]{thom08}.

Still, questions about HVC origins and their role in the evolution of the Milky Way remain.  For example, HVCs have been posited ever since their discovery as a potential source of star formation fuel for the Milky Way via gas accretion onto the disk.  However, an estimate of the total mass ($M$) delivered by HVCs is severely hampered by our ignorance of their distance (as $M \propto d^2$). Conversely, pinpointing their distance would not only allow an account of their mass budget, but would also render HVCs useful test probes of the poorly constrained Milky Way's hot halo.  

The question about HVC distances goes hand in hand with the mystery about their survivability as they move through the halo. Na\"\i{}vely, the clouds are expected to be destroyed by hydrodynamic (HD) -- i.e. Kelvin-Helmholtz (KH) and Rayleigh-Taylor (RT) -- instabilities within times scales on the order of a few 10 \Myr\ \citep{heitsch09}. However, HVCs must survive for longer than this based on the distances that we do have to some complexes, assuming they did not originate at or close to their present location. A notable example is the Smith Cloud, a massive (M$_{tot} \approx 2 \times 10^6 \Msun$), enriched (Z $\approx 0.5  \Zsun$) gas structure only 8 kpc from the Galactic center and 3 kpc from the Galactic plane \citep{lockman08,nichols09,fox16}. Although it has been suggested that the Smith Cloud may in fact be of Galactic origin \citep[][]{fox16,marasco17}, the difficulties of the proposed models (in particular the energy requirements) make this suggestion rather implausible. Instead, the Smith Cloud is likely an extragalactic system accreted by the Galaxy that has survived its journey all the way to the disk \citep[e.g.][see also \citealt{henley17}]{nichols14}.

It is now recognized that spiral galaxies, including the Milky Way, are surrounded by a magnetized medium (see \citealt{beck16} for a review). While the detailed structure of the Galactic magnetic field is still uncertain, the proposed models more or less agree on the overall shape and strength \citep[e.g.][]{sunreich10,jansson12a}. In addition, the relatively recent discovery of enhanced Faraday rotation measures at the Smith Cloud \citep{hill13} and the Leading Arm \citep[LA;][]{mcclure-griffiths10} indicates that magnetic fields can no longer be ignored in the study of HVCs. The idea that they may provide a means to suppress instabilities, thus prolonging the lifetime of clouds, has been explored in the past using numerical simulations of cloud-wind interactions which include the effect of magnetic fields \citep[e.g.][]{gregori99,dursi08,banda-barragan16,mccourt15,goldsmith16}. However, these studies are inconclusive as they have yielded mixed results, indicating that magnetic fields may either strongly \citep{mccourt15,goldsmith16} or only mildly \citep{banda-barragan16} suppress HD instabilities, or even enhance these \citep{gregori99,gregori00}. It should be noted that \cite{goldsmith16} actually simulated a cloud-shock interaction rather than a cloud-wind interaction. While these two scenarios are usually thought to be closely related, the recent study of \cite{goldsmith17} suggests that there can be significant differences in their evolution.\\

In this paper, we use magnetohydrodynamic (MHD) simulations to explore the evolution of HVCs as they move through the magnetized, hot, diffuse Galactic halo toward the Galactic plane. We focus on the interaction of the gas with the magnetic field at the cloud-halo interface, taking into account the variation in density and field strength along the cloud's orbit. We argue that the resulting field amplification provides a robust, though coarse, constraint on the cloud's distance. In passing, we briefly address the effect of the magnetic field on the survival of the cloud. A more detailed study of cloud survival over an extended parameter space of different cloud properties and magnetic field configurations will be reported elsewhere.


\section{Numerical experiments}
\label{sect:ICs}
The most intuitive and straightforward approach to simulate an extragalactic gas cloud falling toward the Galactic disk is to initialize a domain with a realistic galaxy halo model, at least spanning the full distance from the cloud's position, say $d \sim 50 \kpc$, to the Galactic plane with the appropriate density and magnetic field strength gradients. However, both the Galactic gas density and its magnetic field vary by several orders of magnitude across such a galactocentric distance \citep{tepper-garcia15,sunreich10,jansson12a}, and the computational requirements imposed by this large dynamic range are  demanding. We opt for an alternative approach, which allows us to carry out realistic simulations at sufficient resolution and within reasonable computation times. 

In essence, we perform the following {\em Gedankenexperiment}: Consider a relatively small, rectangular domain with open boundaries, enclosing a dense gas cloud at a given position in the Galactic halo. Now trace a desired orbit from that point to the Galactic plane, and let the box move at constant velocity along this orbit with the cloud initially being comoving. As the box moves through the halo, the cloud will experience drag augmented by the increase in density and the gradient in the ambient magnetic field, until it eventually reaches the disk.
We start in the rest frame of the cloud but rather than accelerating our simulation frame relative to the halo at later times to stay in the cloud's rest frame as the cloud experiences drag we simply continue to move it at the constant initial velocity. Such an approach has been used with success in the past albeit in a different context \citep[e.g.][]{nichols15,salem15}.

In practice, we fill the computational domain with an initially uniform, hot medium (`halo gas') with density $n_h$ and magnetic field $B_h$ appropriate for the orbit's initial point in the halo. Close to the leading boundary we insert an initially spherical cloud, with a smooth density profile described by
\begin{equation} \label{eq:densprofile}
	n(r)=n_h+\frac{n_{c}-n_h}{1+(r/r_{\text{c}})^s} \, ,
\end{equation}
where the subscript `$c$' refers to the cloud. At runtime, a magnetized `wind' of hot, diffuse material is injected through the domain's leading boundary with constant speed $v_{\text{wind}}$, and its properties are varied appropriately in time to consistently account for the density gradient and the changing field strength of the halo along the cloud's orbit. Note that our approach to set the initial halo gas density and the initial magnetic field \emph{uniformly} across the simulation volume is justified, since the cloud remains close to the domain's leading boundary at all times, and the variation of both density and magnetic field between the boundary and the cloud is in fact negligible. 

\begin{figure}[h]
\begin{center}
\includegraphics[width=0.49\textwidth]{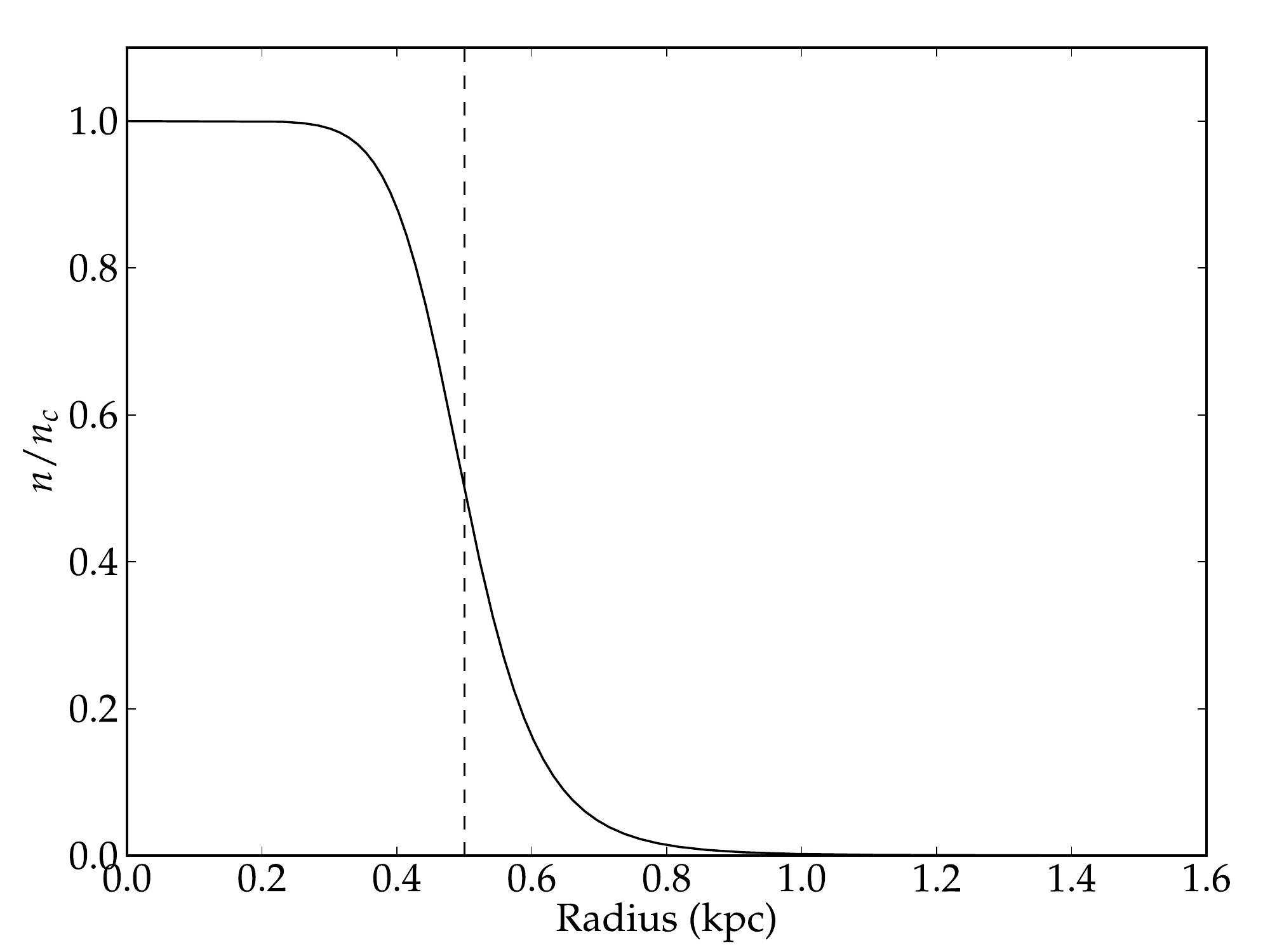}
\end{center}
\caption{Initial density profile of our model HVC. The dashed line shows the cloud radius where the density is approximately half of the central value \citep[cf.][]{banda-barragan16}.}
\label{fig:densprofile}
\end{figure}

To account for the variability in density among HVCs, we consider two representative cases: a low and a high density cloud. The initial density in the cloud core is set to $n_{c}=0.1 \pcc$ ($n_{c}=0.5 \pcc$) for the low (high) density case, yielding an initial cloud mass $M_{\text{c}} \approx 7 \times 10^5 M_\odot$ ($M_{\text{c}} \approx 3 \times 10^6 M_\odot$), the latter being comparable to the mass of e.g. the Smith Cloud \citep{lockman08}. In either case, the cloud radius is set to $r_{\text{c}}=0.5$ kpc, and the parameter $s$ -- which determines the steepness of the profile -- to $s=9$ (see Figure \ref{fig:densprofile}). Note that for $n_{c} \gg n_h$ as is the case in all our simulations, $n(r_{\text{c}}) \approx n_{\text{c}}/2$.

We choose the computational domain to be a uniform, rectangular, 3D grid composed of $256 \times 256 \times 1152$ cells with virtual physical dimensions $8 \kpc \times 8 \kpc \times 36 \kpc$, spanning a coordinate system $(x',y',z')$ initially in the rest frame of the cloud and defined by $x'=(-4, 4)$, $y'=(-4, 4)$ and $z'=(-2, 34)$ for simulations with low density clouds. For simulations with high density clouds the coordinate system spans $8 \kpc \times 8 \kpc \times 28 \kpc$  with $z'=(-2, 26)$ and the grid is proportionally smaller.\footnote{In either case, the domain's leading boundary is at $z = -2$ kpc. The size of the box is chosen such that only an insignificant amount of the material that is ablated from the cloud by the wind leaves the domain before the end of the simulations.}. It is worth noting that the ratio of box dimension to cloud radius translates into a resolution of 16 cells per cloud radius, which is sufficiently high to capture the general evolution of the magnetic field strength (see Section \ref{sect:convergence}).

Positions in the halo are identified by coordinates given in a fixed Cartesian coordinate system $(x,y,z)$ with origin at the Galactic center. In this system, $z$ is the distance from the Galactic plane and $x$ measures the distance of the cloud along the Galactic plane in the direction of the cloud's initial center of mass. Since we assume that the halo density and field structure is axisymmetric (see below), the initial location of the cloud is fully identified by a pair of coordinates denoted by $(x_0, z_0)$. To simplify further, we choose the cloud's orbit to be perpendicular to the Galactic plane at all times. In other words, the cloud's orbit is fully specified by the value of $x_0$.

The above setup implies that the simulation domain `moves' in the negative $z$ direction (starting from $z_0$) with speed $v_{\text{wind}}$ with respect to the fixed frame $(x,y,z)$. Thus, the moving and fixed frame coordinates (in units of kiloparsec) are related through the Galilean transformation
\begin{equation} \label{eqn:coordtransform}
	(x,y,z)=(x'+x_0,y',z'+z_0-  10^{-3} v_{\text{wind}}t) \, ,
\end{equation}
where $v_{\text{wind}}$ is in units of \kms, $t$ is the simulation time in Myr, and the numerical factor (approximately) accounts for the conversion from $\kms$ to kpc Myr$^{-1}$.

\begin{figure}[h]
\begin{center}
\includegraphics[width=0.47\textwidth]{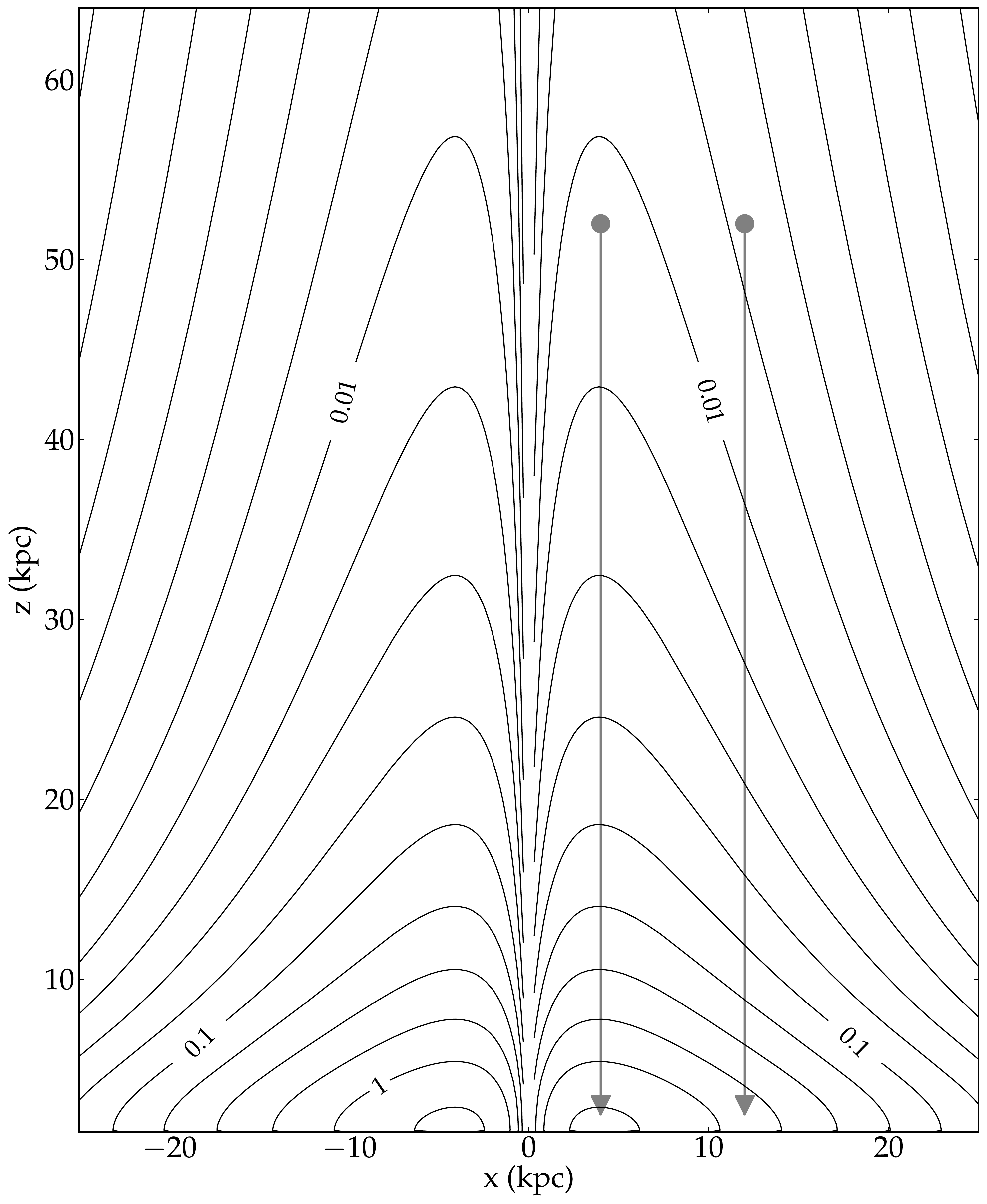}
\end{center}
\caption{The halo magnetic field adopted in our simulations: the toroidal component of the \cite{sunreich10} model. Each contour indicates the field strength on a slice through the $xz$ plane at $y=0$ in galactocentric coordinates. Since the field is purely azimuthal in cylindrical coordinates, the angle of the $xy$ plane about $z$ is arbitrary and the field vector points everywhere either into or out of the page depending on the sign of $x$ (except at $x=0$ where it vanishes). The two initial cloud positions $x_0=4$ kpc and $x_0=12$ kpc are marked with filled dots, their trajectories are marked with arrows.}
\label{fig:bmaghalo}
\end{figure}

The variation in halo density along the cloud's pre-defined orbit is calculated using the spherically symmetric, isothermal, standard model by \cite{tepper-garcia15}, which has been shown to reproduce well the average density profile of the Galactic hot halo out to a radial distance of $r \approx 250 \kpc$. The magnetic field is calculated using the Galactic magnetic field from \cite{sunreich10}. This model has three components: a toroidal (i.e. axisymmetric) halo field, an isotropic random field; and a disk field. We ignore the latter as it is insignificant at $z \gtrsim 1$ kpc (the gas disk density is also ignored for the same reason). We also ignore the random field component because its nature and strength are highly uncertain. Indeed, in the \cite{sunreich10} model the random field has a uniform magnitude throughout the halo, while in the \cite{jansson12a,jansson12b} model the random field is a Gaussian around $z=0$ and is insignificant compared to the halo field for the distances we consider. The halo field's axisymmetric component has a simple analytical form of
\begin{align} \label{eqn:bfield}
B_\phi(R,z)	= & \frac{\text{sign}(z)B_0}{1+\left\{ (\vert z \vert-z_a) / z_b \right\}^2} \notag \\
& \times \left( \frac{R}{R_0} \right) \exp{\left[-\frac{R-R_0}{R_0}\right]}
\end{align}
in cylindrical coordinates where $z_a=1.5$ kpc, $z_b=4$ kpc, $B_0=2\mu$G and $R_0=4$ kpc (see Figure \ref{fig:bmaghalo}).

The initial magnetic field is (see equation \ref{eqn:bfield})
\begin{equation}
\label{eqn:bhalo}
\vec{B} = (B_x,B_y,B_z)=(0,B_\phi,0),
\end{equation}
i.e. initially the field is uniform and points in the positive $y$ direction throughout the entire volume. In this way we are ignoring the variation in direction of the field across the $xy$ plane as well as changes in its magnitude. Because it has no vertical component, it is everywhere transverse to the cloud's orbit as defined above.

\begin{figure}[h]
\begin{center}
\includegraphics[width=0.49\textwidth]{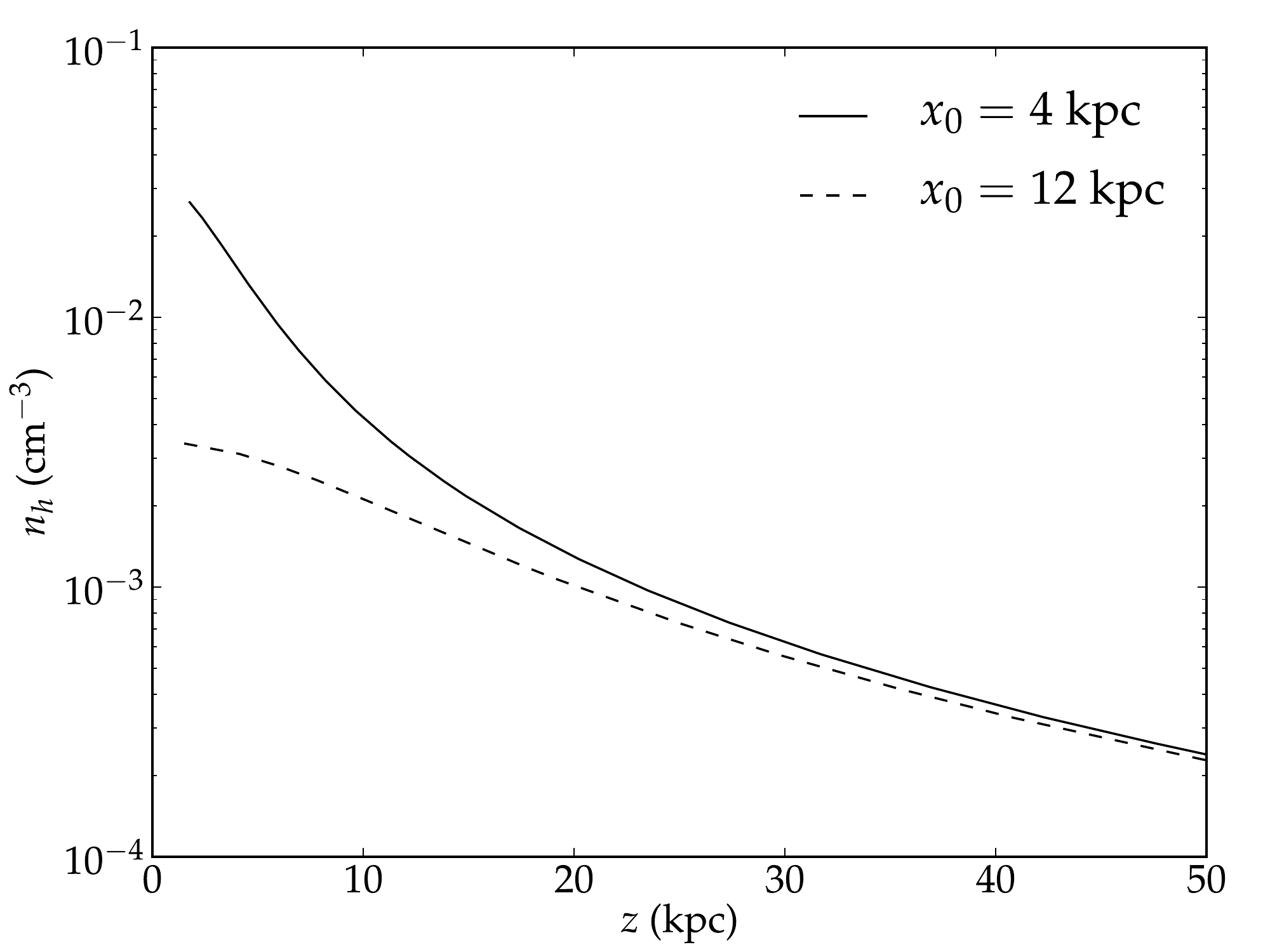}
\includegraphics[width=0.49\textwidth]{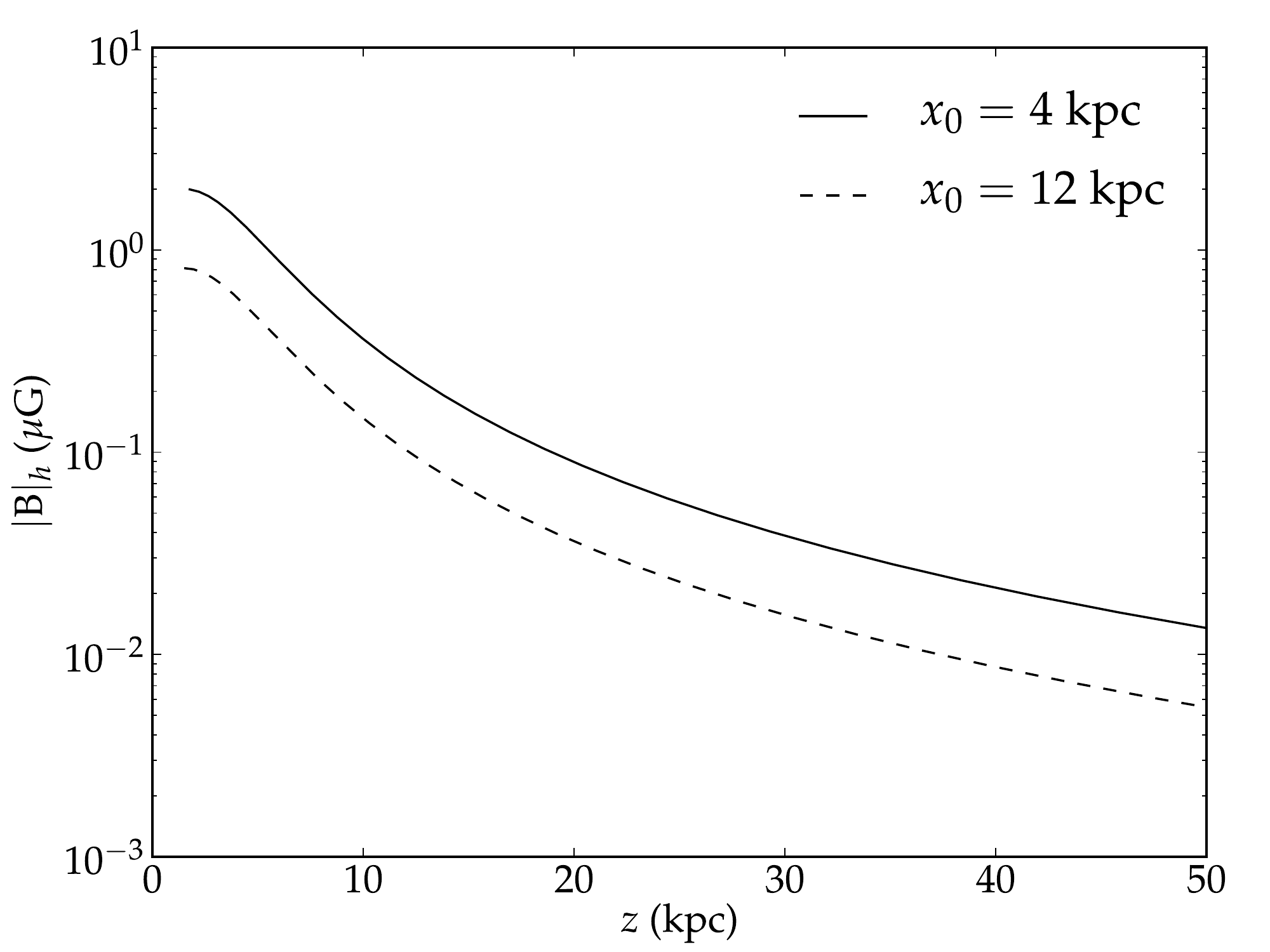}
\end{center}
\caption{Halo density (top) and magnetic field strength (bottom) along an HVC's orbit starting at $z_0 = 52 \kpc$ and $x_0=4 \kpc$ (solid) or $x_0=12 \kpc$ (dashed). The Galactic plane is at $z = 0 \kpc$.}
\label{fig:haloprofiles}
\end{figure}

Densities are linearly interpolated in $z$ using a table based on the \cite{tepper-garcia15} standard model with a resolution that decreases with distance from $\Delta z=0.2$ to $\Delta z = 2.4$ kpc. Magnetic field strengths are calculated for each $z$ from equation \eqref{eqn:bhalo} which in this case is just $\vert B\vert \propto 1/(1+[(z-z_a)/z_b]^2)$. The halo density and magnetic field strength at $x_0=4$ kpc and $x_0=12$ kpc as function of $z$ along the cloud's orbit are shown in Figure \ref{fig:haloprofiles}.

To simulate the motion of the HVC (i.e. of the box) along its orbit, we calculate the distance from the Galactic plane, $z$, of the domain's leading boundary at each time step through equation \eqref{eqn:coordtransform} and set the density and magnetic field there accordingly. The velocity there is set to be $-v_{\text{wind}}$ throughout.  At the start, both the density and the magnetic field are set to their values at the initial position of the leading boundary at $z = 50 \kpc$ close to the cloud's initial position of $z_0 = 52 \kpc$. These are $n \approx 2 \times 10^{-4} \pcc$ and $\vert B\vert \approx 0.03 \mG$ at $x_0=4$ kpc and $n \approx 2 \times 10^{-4} \pcc$ and $\vert B\vert \approx 0.01 \mG$ at $x_0=12$ kpc. Note that the density is roughly equal at either point because it has a spherically symmetric profile and $x_0 \ll z_0$ so the distance from the Galactic center to the cloud is approximately the same in either case. Initially, all the material outside the cloud and the cloud-halo transition region defined as $r>2r_c$, is set have velocity $-v_{\text{wind}}$ such that the halo material is stationary in the Galactic coordinates. We neglect the variation of the halo magnetic field and density across the simulation volume transverse to the cloud's orbit, i.e. we set these quantities to be equal to their value at the cloud's position, $x=x_0,y=0$, for all $x$ and $y$ at the injection boundary. Including this radial variation in the simulations is straightforward but we choose not to do so in order to have a one-to-one relation between $z$, density and magnetic field strength in the halo. That said, we did run a set of simulations that include the radial variation of the density and the magnetic field to check the validity of ignoring the radial variation in other simulations.  For the magnetic field this also entails that, as in the initial conditions, it continues to point in the $y$ direction throughout the volume rather than circling around the $z$ axis. It would be entirely transverse to the cloud in either case, however. We have verified that ignoring the radial variation is a valid approximation as discussed in Section \ref{sect:limitations}. Needless to say, this approach greatly simplifies the analysis, while keeping our simulations realistic enough.\\

We calculate the time evolution of an HVC by solving the ideal magnetohydrodynamic (MHD) equations\footnote{The source code is available upon request from the corresponding author}. For this purpose, we use the code PLUTO 4.1 \citep{mignone07,mignone12}. We show that the ideal MHD approximation is appropriate for our simulations in Section \ref{sect:limitations}. We employ the \emph{Constrained Transport} (CT) scheme \citep[][]{evans88} to ensure that the initial divergence of the magnetic field is maintained during the course of the simulation. Indeed,  unlike other schemes that deal with magnetic field divergence such as hyperbolic divergence cleaning, CT does not minimize the divergence but rather keeps it constant in time to within machine precision. Because the initial magnetic field is uniform $\nabla \cdot \mathbf{B}=0$ to machine precision in the initial conditions.

Throughout a run, the halo is assumed to be isobaric, and thus the pressure is held constant at the leading boundary. This implies that the halo gas temperature changes with distance as $T(z)\propto n(z)^{-1}$. For a halo temperature on the order of $10^6 \K$ far away from the Galactic plane, the resulting temperature at $z=1.5$ kpc is $T \sim 10^3 - 10^4 \K$. Note that this is not fully consistent with the \cite{tepper-garcia15} halo model, assumed to be isothermal. However, their halo density profile is consistent with the profile obtained in more elaborated, non-isothermal models \citep{faerman17}. In addition, we require the HVC gas to be initially in pressure equilibrium with the halo. In doing so, we neglect the magnetic pressure $P_B = | \vec{B} |^2 / 8 \pi$, which is justified since it is initially very weak compared to the gas pressure, i.e. $\beta \equiv P_c / P_B \gg 1$. We ignore radiative cooling and photo-heating, and adopt an adiabatic equation of state with index $\gamma=5/3$, appropriate for a monoatomic gas. With the exception of the leading boundary, outflow boundary conditions are imposed everywhere, implying that material is free to flow out of (but not into) the simulation volume.\\

We ran a total of 16 simulations with different initial conditions, varying the value of a single parameter from run to run. The parameters being varied are the density of the cloud, $n_c$; the cloud's initial velocity, equal in magnitude but opposite in sign to $v_{\text{wind}}$; and the initial $x$ position of the cloud, $x_0$. Table \ref{table:ics} lists our runs as well as the parameter values adopted in each one. Note that the parameter values are representative of Galactic HVCs. For simplicity, we only adopt two different values for each of the relevant parameters. In addition, we run three simulations with no magnetic field, two simulations with higher resolution, and three simulations where the radial variation across the simulation volume of the density and magnetic field is included. The purpose of the simulations without magnetic fields is to assess the impact of magnetic fields on cloud survival (see Section \ref{sect:survival}). The high resolution simulations are used for a crude convergence test (see Section \ref{sect:convergence}). All simulations are run up to the point where either the cloud's center of mass reaches $z \approx 1.5$ kpc, or the cloud is slowed by drag in the Galactic rest frame essentially becoming comoving with the wind. Note that in all cases the simulation domain is large enough to avoid that more than a few per cent of the initial cloud mass has left the simulation domain by the end of the run. In order to track the cloud's evolution more accurately, we tag all cells initially within the cloud with a passive scalar. In terms of dimensionless quantities, the initial cloud-halo density contrasts and wind Mach numbers for $v_{\text{wind}}=200 \kms$ are $\chi \approx 500$ and $\mathcal{M} \approx 3.5$ for $n_c=0.1 \pcc$ and $\chi \approx 2500$ and $\mathcal{M} \approx 1.5$ for $n_c=0.5 \pcc$, respectively. Mach numbers for $v_{\text{wind}}=300 \kms$ are obtained from the former by scaling them up by a factor of 1.5. Our simulations hence fall in the transonic to supersonic regime. Note that the Mach number in each case depends on the cloud density because of our initial isobaric conditions everywhere in the volume. Neither case depends noticeably on $x_0$ because the halo density depends on distance to the Galactic center which is approximately the same for $x_0=4$ kpc and $x_0=12$ kpc at the initial height above the plane of $z_0=52$ kpc.  

\begin{deluxetable}{lcccc}
\tablewidth{\textwidth} 

\tablecaption{Model parameters \label{table:ics}}

\tablehead{\colhead{Name$^a$} & \colhead{$n_c$ $^e$} & \colhead{$v_{\text{wind}}$} & \colhead{$x_0$} & \colhead{Resolution$^f$}\\ 
\colhead{} & \colhead{( \pcc)} & \colhead{(\kms)} & \colhead{(kpc)} & \colhead{(cells/$r_c$)} } 

\startdata
L200x4    & 0.1 & 200 & 4  & 16\\
H200x4    & 0.5 & 200 & 4  & 16\\
L200x12    & 0.1 & 200 & 12 & 16\\
H200x12    & 0.5 & 200 & 12 & 16\\
L300x4    & 0.1 & 300 & 4  & 16\\
H300x4    & 0.5 & 300 & 4  & 16\\
L300x12    & 0.1 & 300 & 12 & 16\\
H300x12    & 0.5 & 300 & 12 & 16\\
L200x4-HD$^b$ & 0.1 & 200 & 4  & 16\\
H200x4-HD$^b$ & 0.5 & 200 & 4  & 16\\
L300x4-HD$^b$ & 0.1 & 300 & 4  & 16\\
L200x4-medres$^c$ & 0.1 & 200 & 4  & 24\\
L200x4-hires$^c$ & 0.1 & 200 & 4  & 32\\
L200x4-rv$^d$ & 0.1 & 200 & 4  & 16\\
H200x4-rv$^d$ & 0.5 & 200 & 4  & 16\\
L200x12-rv$^d$ & 0.1 & 200 & 12  & 16\\
\enddata
\tablecomments{$^a$ The naming convention is as follows: The first letter, L (H), indicates a run adopting a low (high) density HVC; the following three-digit number indicates the adopted velocity in \kms; the number following `x' corresponds to the cloud's $x$ position, $x_0$, in kpc. $^b$ This run does not include a magnetic field, but is otherwise identical to its MHD counterpart. $^c$ This run adopts a higher resolution (see last column). $^d$ This run takes into account the radial variation of the halo density and magnetic field along $x$ and $y$ across the simulation volume. $^e$ Initial density contrasts are $\chi\approx 500$ for $n_c=0.1 \pcc$ and $\chi\approx 2500$ for $n_c=0.5 \pcc$ $^f$ Given as the number of cells per cloud radius.}
\end{deluxetable}

\section{Results} \label{sec:results}

The evolution of physical quantities is followed in terms of the distance traveled by the cloud's center of mass (CoM) given by $z_0 - z_{\text{\sc cm}}$ rather than its distance from the Galactic plane, $z_{\text{\sc cm}}$. In all simulations the initial distance from the Galactic plane is $z_0=52$ kpc. In addition to the figures that we show in this section, animations, including 3D renderings, are available at \url{http://www.physics.usyd.edu.au/~agro5109/animations.html}.

\subsection{Cloud distances}

In this section, we will restrict our discussion to simulations L200x4, H200x4 and L200x12, all of which correspond to a cloud initially moving at 200 \kms, but with different initial density and initial position. This is because we found no significant differences with distance between the clouds initially moving at 200 \kms\ and 300 \kms.\footnote{It is worth emphasizing that there {\em is} a difference between the evolution of the $v_{\text{wind}}=200 \kms$ and the $v_{\text{wind}}=300 \kms$ simulations in terms of time, but this difference becomes irrelevant when time is rescaled by the initial velocity ratio.}

\begin{figure}[h]
\begin{center}
\includegraphics[width=0.49\textwidth]{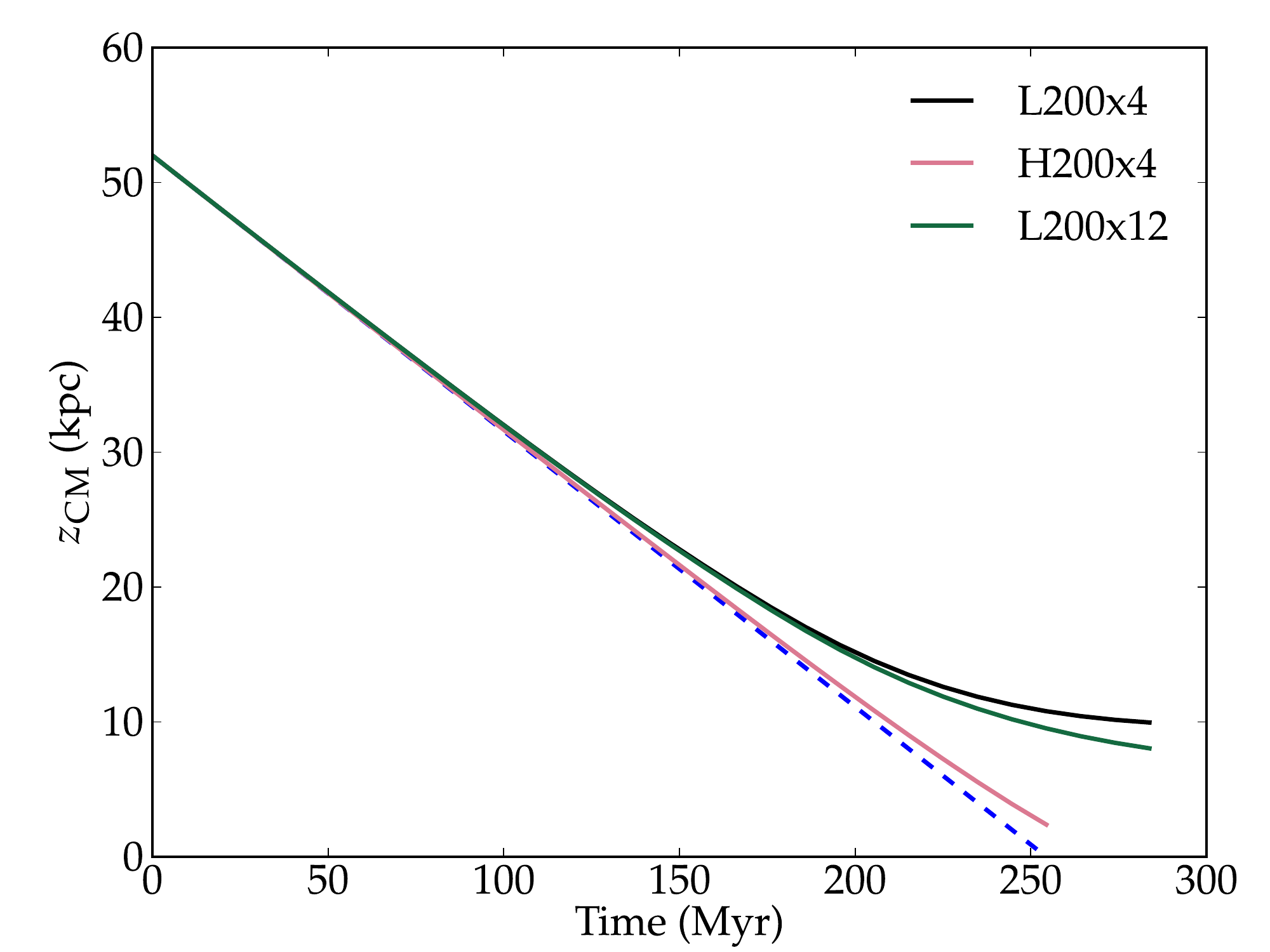}
\end{center}
\caption{Distance from the Galactic plane of the HVCs' center of mass as function of time, initially moving at 200 \kms. The dashed line corresponds to motion at constant velocity, i.e. how the distance to the clouds would evolve in the absence of drag. The results for clouds initially moving at 300 \kms are similar when rescaled by 1.5, and are therefore omitted.}
\label{fig:timedist}
\end{figure}

Figure \ref{fig:timedist} shows the position of the cloud's center of mass along its orbit ($z_{\text{\sc cm}}$) as a function of time. The dashed line corresponds to motion at a constant velocity of 200 km s$^{-1}$ toward the plane, and is included for reference. Clearly, the clouds move at essentially constant velocity along a large fraction of their orbit in all cases. Low density clouds are eventually hampered by hydrodynamic and magnetic drag, which become significant at $z \sim 15 \kpc$, efficiently decelerating the cloud to the point that it becomes comoving with the wind, i.e. stationary in $z$, at $z \approx 10$ kpc. Clouds falling further away from the center at $x_0=12 \kpc$ are able to travel slightly larger distances due to the weaker drag resulting from a lower halo density at larger radii. High density clouds move nearly unimpeded with a velocity close to its initial value all the way to the smallest allowed distance of $z = 1.5 \kpc$, potentially reaching the disk. Note that in all cases, the motion of the cloud is (highly) supersonic, with Mach numbers in the range of $\sim 2 -  12$.\footnote{Approximate Mach numbers for the $v_{\text{wind}}=300$ km s$^{-1}$ case are obtained by scaling by a factor 1.5.}

\subsection{Magnetic field amplification}

The fundamental difference between HD and ideal MHD is the presence in the latter formalism of the induction equation,
\begin{equation} \label{eqn:induction}
	\frac{ \partial \vec{B} }{ \partial t } = \vec{\nabla} \times \left( \vec{v} \times \vec{B} \right) \, ,	
\end{equation}
which describes the evolution of a magnetic field embedded in a fluid in motion. In essence, the induction equation is a statement of the fact that a magnetic field moving with a fluid, no matter how small and regardless of the details of the motion, will eventually be amplified, potentially by factors of several orders of magnitude. In numerical experiments like ours, the clouds move through a weakly magnetized medium and `sweep up' the ambient field along their orbit, compressing and amplifying it along the direction of motion at their leading edge and in the transverse direction in their wake \citep[see also e.g.][]{dursi08}. Motivated by the fact that enhanced fields have been observationally associated with some of the Galactic HVCs (see Sec. \ref{sec:intro}), we perform in the following a detailed analysis of the evolution of the magnetic field in our simulations, especially paying attention to its behavior around the cloud.

The importance of the magnetodynamic effects relative to hydrodynamical effects can be measured by the ratio of gas pressure to magnetic pressure $\beta = 8 \pi P / B^2$. In our simulations, $\beta \approx 2 \times 10^3$ everywhere initially, implying that the magnetic field is dynamically irrelevant. Note that our assumption of an isobaric halo ($P$ fixed) implies $\beta \propto B^{-2} \propto (1+[ (z-z_a) / z_b ]^2)^2$ in the halo. Thus, for $\chi=2500$ and at $x_0=4$ kpc, where the the constant of proportionality is $8 \pi P B_0^{-2} \approx 0.1$, $\beta<1$ for $z<8$ kpc, i.e. close to the plane. In general, the value of $\beta$ anywhere in the volume remains well above unity throughout the majority of the evolution of the clouds, regardless of their initial density or velocity, and thus hydrodynamic effects dominate overall. Nevertheless, magnetodynamic effects rapidly gain importance with time, as indicated by the evolution of $\beta$, which reaches $\beta \approx 0.4$ in simulation H200x4 and $\beta \approx 0.07$ in simulation L200x4 for all clouds near the Galactic plane, at the 10th percentile level over the whole volume. The relevance of magnetodynamic effects is attested as well by the ratio of ram pressure to magnetic pressure, $8\pi \rho_h v^2 / B^2$ ($\rho_h$ being the mass density of the halo at the leading edge of the cloud). This decreases by several orders of magnitude from an initial value of order $10^4$ to $\sim 10$ for \Bnine\  by the end of simulation H200x4 (and vanishes by the end of the low density cloud simulations as they become comoving with the wind).

\begin{figure}[h]
\begin{center}
\includegraphics[width=0.49\textwidth]{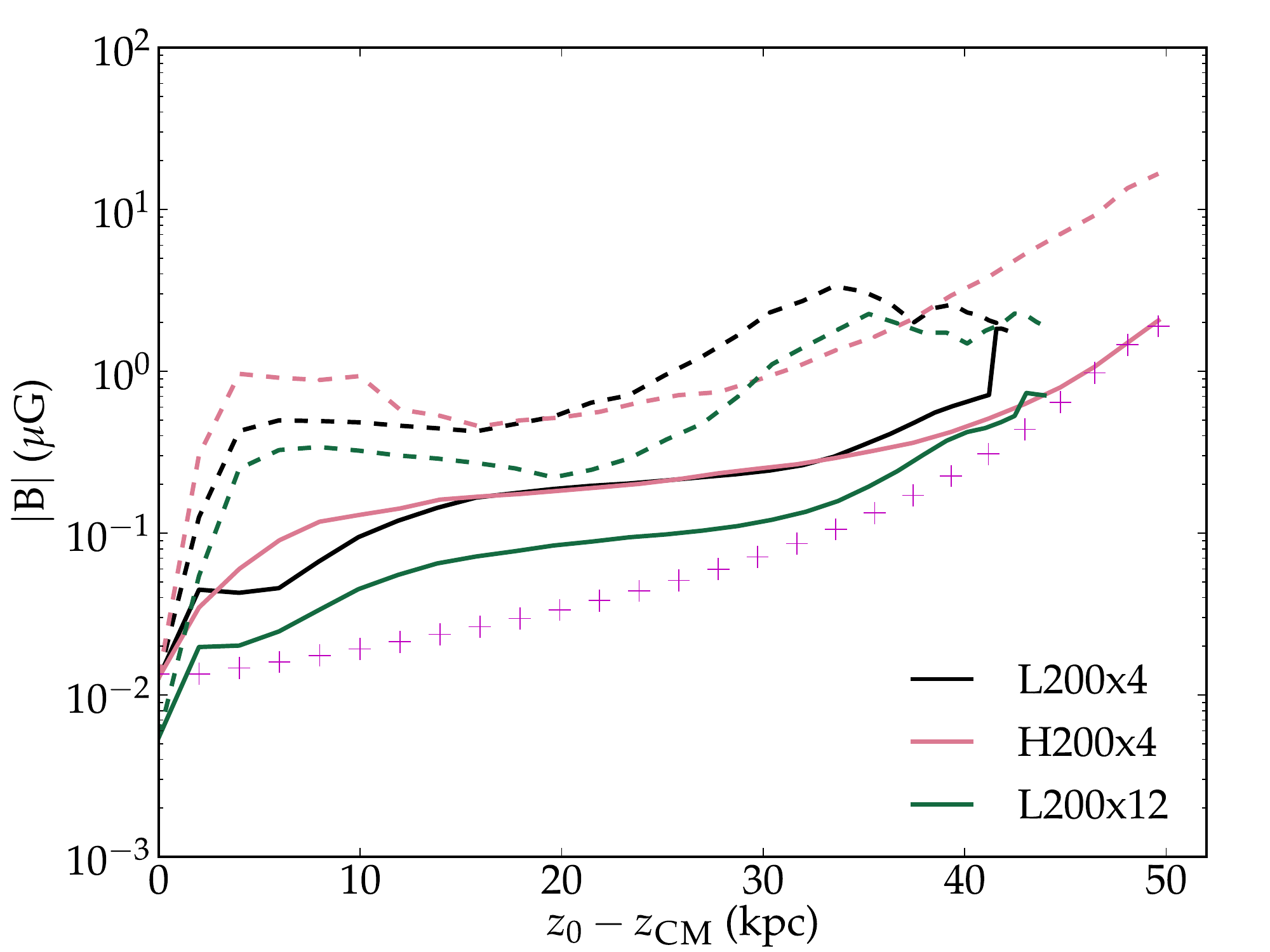}
\end{center}
\caption{The evolution of the overall magnetic field strength along the HVC's orbit. The horizontal axis is the distance travelled by the cloud from $z_0=52$ kpc with $z_{\sc CM}$ being the distance along $z$ of the cloud's center of mass. Solid lines correspond the 90th percentile magnetic field strengths after values close to the halo field have been filtered out, \Bnine\ (see Section \ref{sec:results}), and dotted lines are the maximum field strengths, \Bmax\ . The crosses are the halo field strength at the corresponding $z_{\text{\sc cm}}$ for $x=4$ kpc.}
\label{fig:bmag}
\end{figure}

The evolution of the magnetic field strength in the volume, $\vert B\vert$, as function of the cloud's position along its orbit is shown in Figure \ref{fig:bmag}. For ease of discussion, in what follows we characterize the magnetic field strength $\vert B\vert$ using essentially two quantities: (1) the maximum value of $\vert B\vert$ in the simulation at a given $z_{\text{\sc cm}}$ (i.e. time) denoted by  \Bmax and (2) the 90th percentile of the distribution of magnetic field strength across the volume denoted by \Bnine.  The latter is calculated from the distribution which results after removing all cells within one standard deviation of the local halo field. This approach removes outliers and makes \Bnine\ a robust statistic. In contrast, \Bmax\ may be subject to strong local and temporal fluctuations, but is still useful as an absolute upper limit to the magnetic field strength at any given time.

In Figure  \ref{fig:bmag}, solid (dashed) lines correspond to \Bnine\  (\Bmax) . For low density clouds, \Bnine\  keeps increasing throughout their journey, while the growth of \Bmax\  levels off as the cloud slows down in the wind frame near the Galactic plane and then decays slightly. In contrast, high density clouds experience an ongoing field growth in both \Bnine\  and \Bmax\  all the way to the disk-halo interface.

\begin{figure*}[htb!]
\begin{center}
\includegraphics[width=\textwidth]{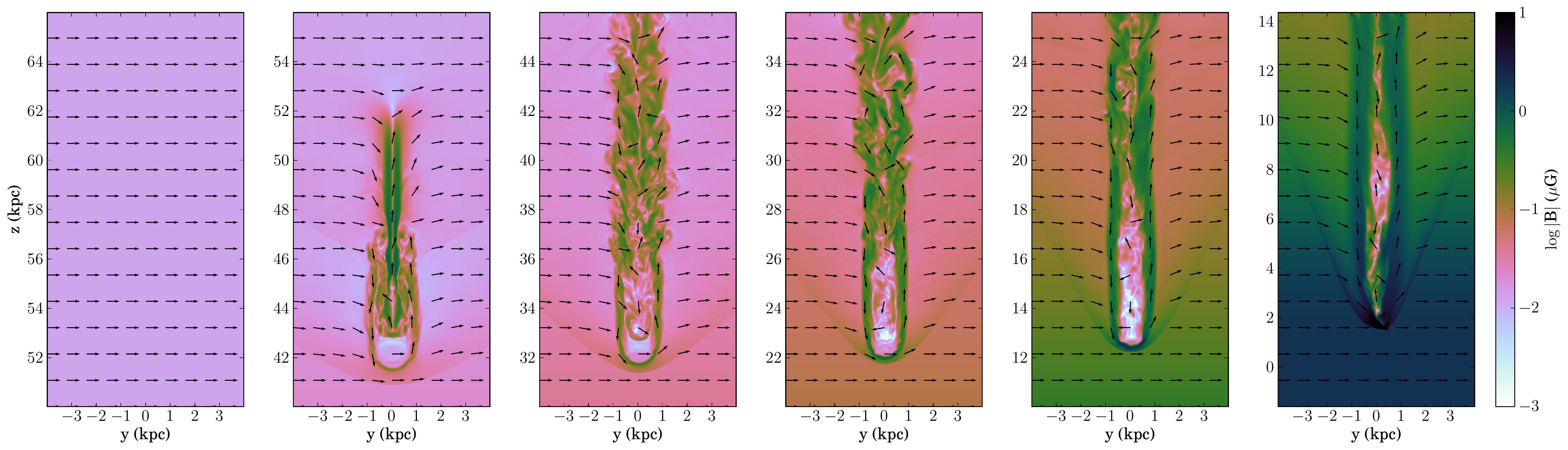}
\end{center}
\caption{Magnetic field strength on a slice through \mbox{$x = 4 \kpc$} in simulation H200x4 at \mbox{$t \approx 0 \Myr$}, 50 \Myr, 100 \Myr, 150 \Myr, 200 \Myr, and 250 \Myr, from left to right. The arrows in each panel show the direction of the magnetic field in this plane. The color coding indicates the value of the magnetic field strength on a logarithmic scale. Note that the $z$-range varies across panels, from highest altitude on the left to lower altitude on the right, such that the cloud's center of mass stays close to the bottom. An animated version of this figure is available in the online edition of the journal.}
\label{fig:2Dbmag}
\end{figure*}

In order to get a sense of the spatial distribution of the amplified magnetic field, and to make a connection with observations, we show in Figure \ref{fig:2Dbmag} the evolution of $\vert B\vert$ on a slice through the $yz$ plane at $x = 4 \kpc$  in simulation H200x4. Interestingly, there are two regimes where the field becomes amplified: (1) at the leading edge of the cloud, where the field is amplified rather quickly and continues to grow at all times and (2) behind the cloud, where an enhanced field of comparable magnitude develops along a planar, coherent double tail. The magnetic field there moves in opposite directions along the tail components thus creating a current sheet where the magnetic field annihilates, a feature typical of MHD flow around a sphere \citep[cf.][]{romanelli14,banda-barragan16}. At $t \approx 100 \Myr$, the magnetic tail loses coherence and becomes turbulent, but it becomes strongly collimated again closer to the Galactic plane as a result of the increased magnetic field there (see Section \ref{sect:survival}). The amplified field at the leading edge of the cloud completely dominates over the the tail field for $t \gtrsim 150$ Myr.

\begin{figure}[h]
\begin{center}
\includegraphics[width=0.49\textwidth]{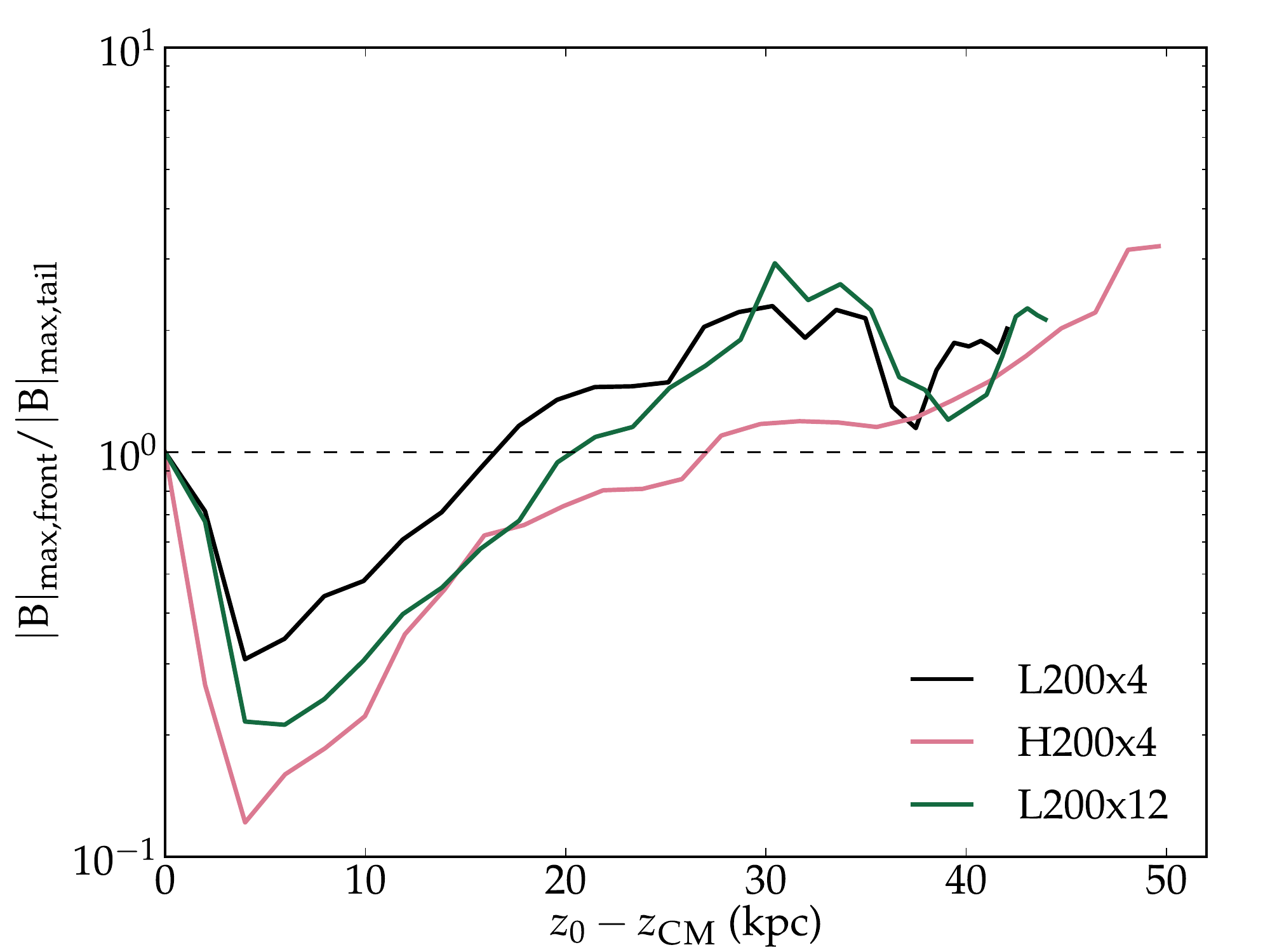}
\end{center}
\caption{Evolution of \Bmax\ at the cloud's leading edge relative to its tail.}
\label{fig:headvstail}
\end{figure}

The relative strengths can be appreciated more quantitatively in Figure \ref{fig:headvstail}. Here we show the evolution of the ratio between the maximum value of $\vert B\vert$ on the leading edge and the \Bmax\ in the tail, defined as $z_{\text{front}} < z_{\text{\sc cm}}$ and $z_{\text{tail}} > z_{\text{\sc cm}}$, respectively. Generally, the tail field dominates at early times while the field at the leading edge dominates at late times. Thus, the front-to-tail field ratio provides some history on the cloud's past interaction. This could be useful in future high-resolution observations that would be able to separately measure the front and tail fields associated with an HVC.

Higher density clouds moving further away from the Galactic center develop stronger tail fields that dominate over the leading amplified field across larger distances. In all cases, the tail field is rapidly amplified at early times only to drop off as the tail becomes elongated and loses its coherence, eventually being overtaken in strength by the field at the leading edge. The leading edge field grows steadily along most of the cloud's orbit,  with a slight decrease at late times for low density clouds. This is a consequence of the stronger drag operating on these clouds close the plane, which efficiently decelerates the clouds and leads to a decrease in the amount of field lines being swept up.

\begin{figure}
\begin{center}
\includegraphics[width=0.49\textwidth]{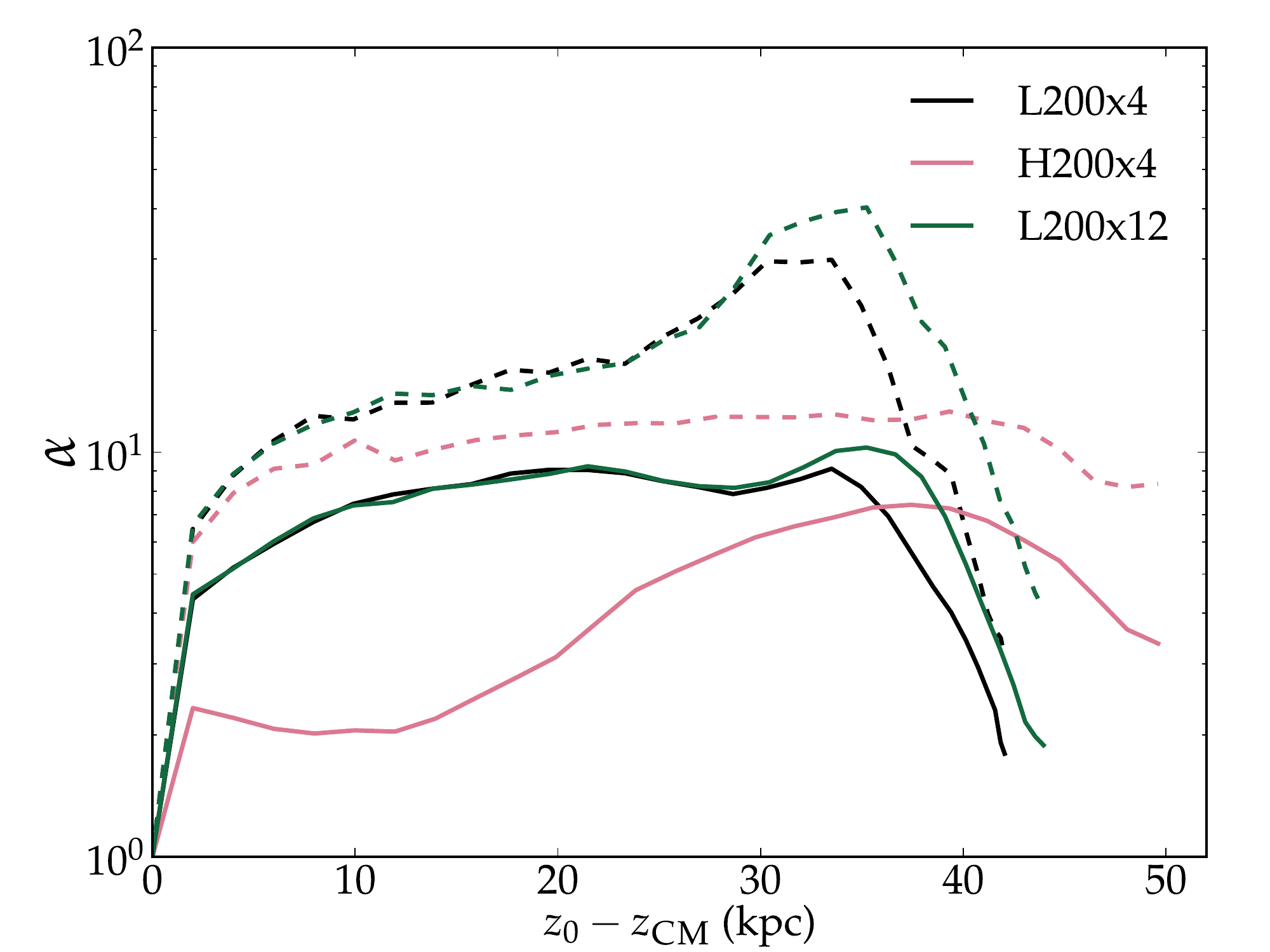}
\end{center}
\caption{Amplification of the magnetic field at the cloud's leading edge relative to the {\em local} halo field, $\alpha \equiv \vert B \vert_{\text{front}} / \vert B \vert_{\text{halo}}$. 
Solid lines correspond to the amplification of $\vert B\vert_{90,\text{front}}$ and dashed lines correspond to the amplification of $B\vert_{\text{max,front}}$.}
\label{fig:bampvshalo}
\end{figure}

Guided by the few available measurements of magnetic fields associated with HVCs so far \citep[][]{mcclure-griffiths10,hill13}, we now consider to be relevant only simulations in which the field: (1) has been amplified beyond the local halo field and (2) is comparable to the observed fields of order 1 \mG. The former is motivated by the fact that the field around the cloud will always be at least equal to the local value. In order to quantify the importance of the field amplification, we consider the ratio of the amplified field to the local field quantified by the parameter $\alpha=\vert$B$\vert/\vert$B$\vert_{\text{halo}}$. In general, $\alpha$ is always highest in the cloud's tail, as material stripped from the cloud which carries magnetic field may be present out to large distances from the Galactic plane where the halo field is much weaker. Therefore we focus on the more interesting case of amplification in front of the cloud. We define $\alpha_{\text{max}}$ as the amplification factor of $\vert B\vert_{\text{max,front}}$ at $z<z_{\text{\sc cm}}$ and $\alpha_{90}$ as the amplification of $\vert B\vert_{90,\text{front}}$  at $z<z_{\text{\sc cm}}$. Here the subscript `front' refers to $z<z_{\sc CM}$, i.e. ahead of the cloud's motion. The evolution of these quantities is presented in Figure \ref{fig:bampvshalo}. Note that this is not generally the same as the maximum and 90 percentile amplification in front of the cloud. We choose this definition because we are only interested in amplification that actually leads to strong fields.
\emph{Low density HVCs experience a more significant field amplification compared to higher density HVCs, moving at roughly the same speed.} However, high-density clouds carry an amplified field at their leading edge for larger distances, i.e. closer to the Galactic plane, as drag does not affect their motion as significantly as it affects lower density clouds. In either case, the amplified field at a cloud's leading edge reaches values that are a factor of at least a few higher than the local field at the 90th percentile level (and at least an order of magnitude higher if the maximum amplification is considered) at intermediate distances.

\begin{figure}[h]
\begin{center}
\includegraphics[width=0.49\textwidth]{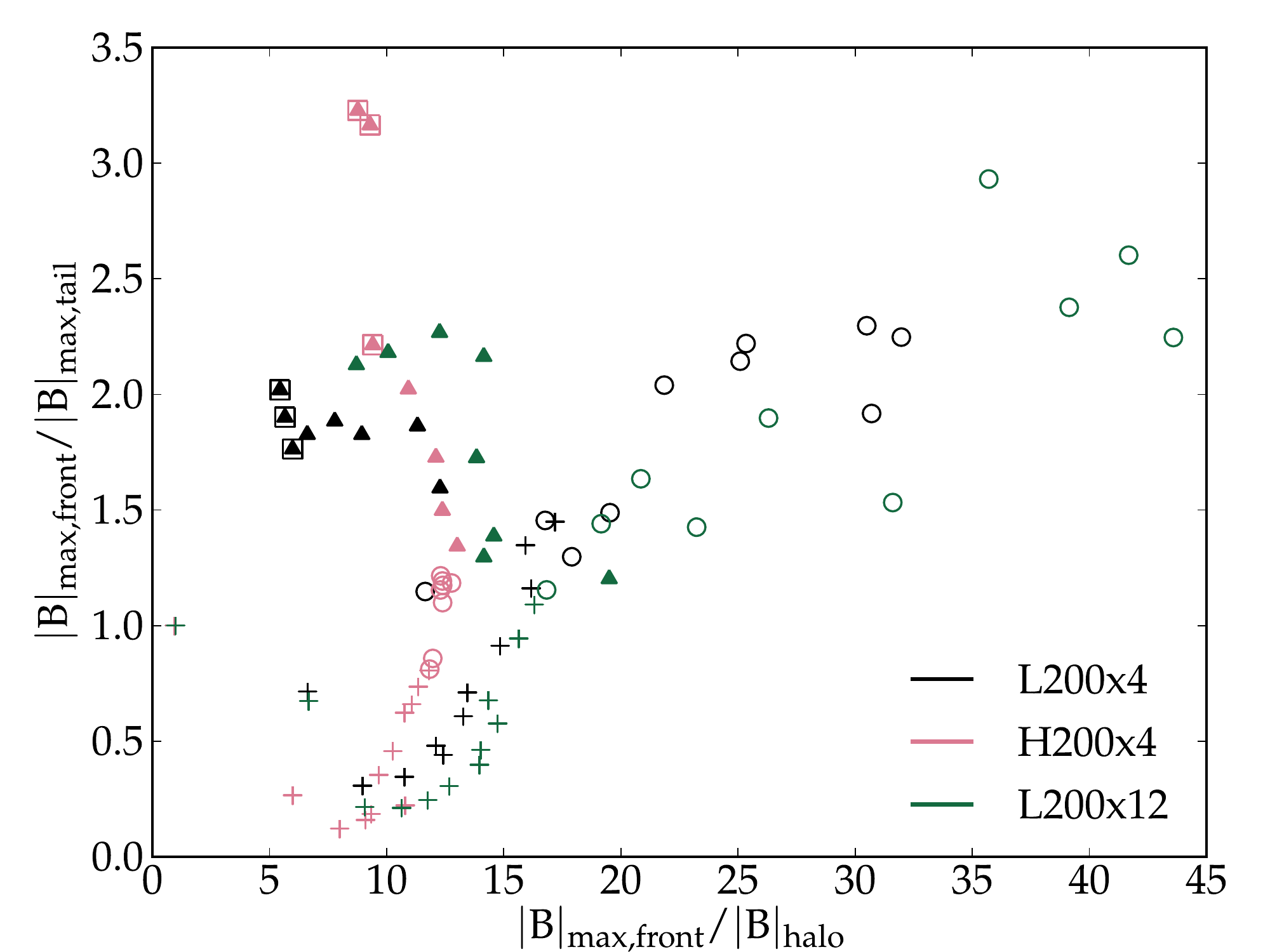}
\end{center}
\caption{Relation of the \Bmax\ at the cloud's leading edge relative to its tail and $\alpha$. Symbols represent different distances from the Galactic plane: Crosses indicate $z_{\sc CM}>30$ kpc, circles indicate $30$ kpc $< z_{\sc CM} < 12$ kpc and triangles indicate $z_{\sc CM} < 12$. Cases where $\Bnine\ > 1\mu$G are enclosed in a square.}
\label{fig:headvstailvshalo}
\end{figure}

In an attempt to condense all our results so far regarding the field amplification, we show in Figure \ref{fig:headvstailvshalo} the relation between the ratio of the field at the cloud's leading edge relative to the tail (q.v. Figure \ref{fig:headvstail}) and $\alpha$ (q.v. Figure \ref{fig:bampvshalo}). In addition, we split the data points in three categories depending on the distance of the cloud's center of mass relative to the Galactic plane. As can be seen the low density clouds trace a path through this plane where they move from a tail dominated phase with maximum amplification of $\alpha_{\text{max}} \sim 10$ to a front dominated phase with $\alpha_{\text{max}}$ reaching $\sim 40$. They then move back down this path and end in a front dominated but relatively low amplification phase. For $x_0=12$ slightly higher amplifications are reached during the intermediate phase. The high density clouds trace out a very different path skipping the intermediate phase of high $\alpha$. Instead, they move mostly vertically in this plane and end up being highly front dominated. In any case, although the greatest amplification occurs at intermediate distances, only relatively close to the plane does the amplified field reach values on the order of \mG.

Thus, taking our results at face value, we conclude that {\em the observed field strengths of $B_{\vert\vert}\gtrsim 5 \mu$G associated with HVCs likely indicate that the clouds must be close to the disk}, i.e. $z \lesssim 10$ kpc. It is worth noting that there is little difference in the results between clouds moving at different velocities and having different initial densities, with exception of the fact that low density clouds hardly reach field values close to $1 \mu$G (ignoring the peak in \Bnine\  at the end of simulation L200x4 which is caused by a numerical artifact). This comes about because -- as previously discussed -- they are unable to reach $z \lesssim 10$ kpc while the high density clouds travel all the way to our minimum allowed distance of $z=1.5$ kpc where the halo field is much stronger.

\subsection{Cloud survival} \label{sect:survival}

\begin{figure}[h]
\begin{center}
\includegraphics[width=0.49\textwidth]{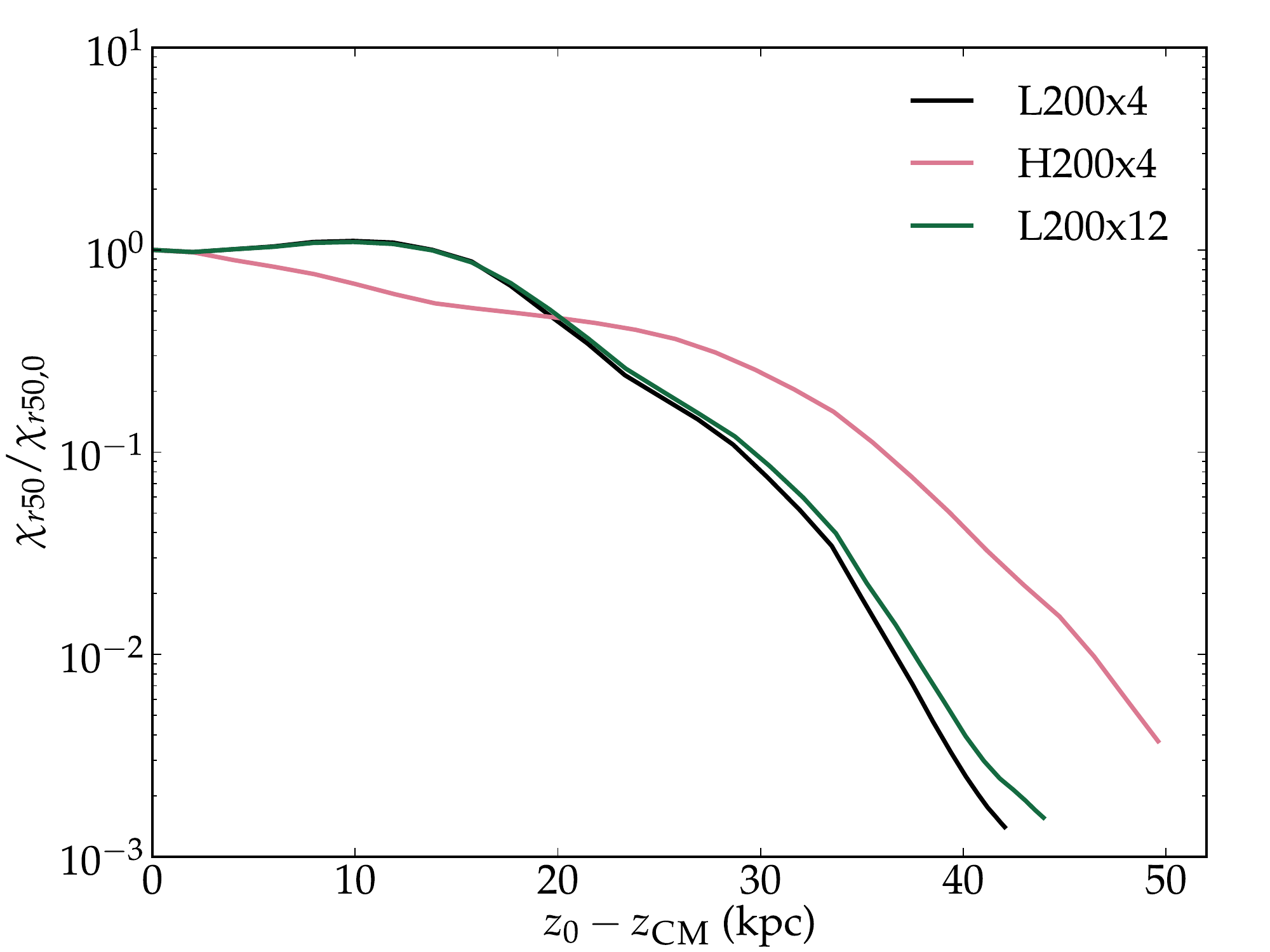}
\end{center}
\caption{Evolution of the mean density of cloud material within the half mass radius $r_{50,\text{sph}}$ relative to the local halo density, normalised by the initial density contrast (see Table \ref{table:ics}).}
\label{fig:denscontrast}
\end{figure}

\begin{figure}[h]
\begin{center}
\includegraphics[width=0.49\textwidth]{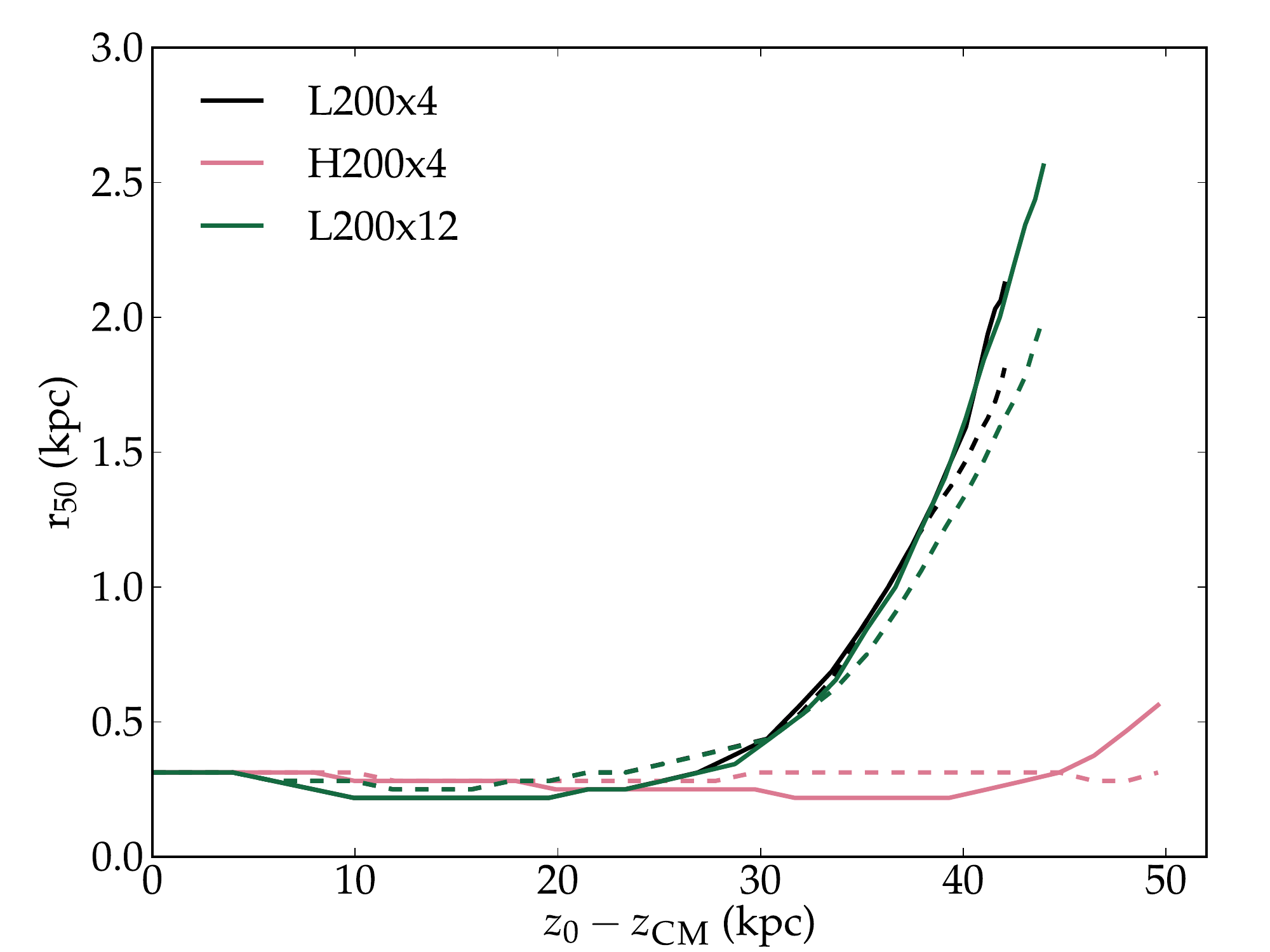}
\end{center}
\caption{Evolution of the cylindrical half mass radius along $x$ (solid lines) and $z$ (dashed lines). The cylindrical half mass radius along $y$ is not shown as it is similar to this quantity along $x$. See text for details.}
\label{fig:r50}
\end{figure}

\begin{figure}[h]
\begin{center}
\includegraphics[width=0.49\textwidth]{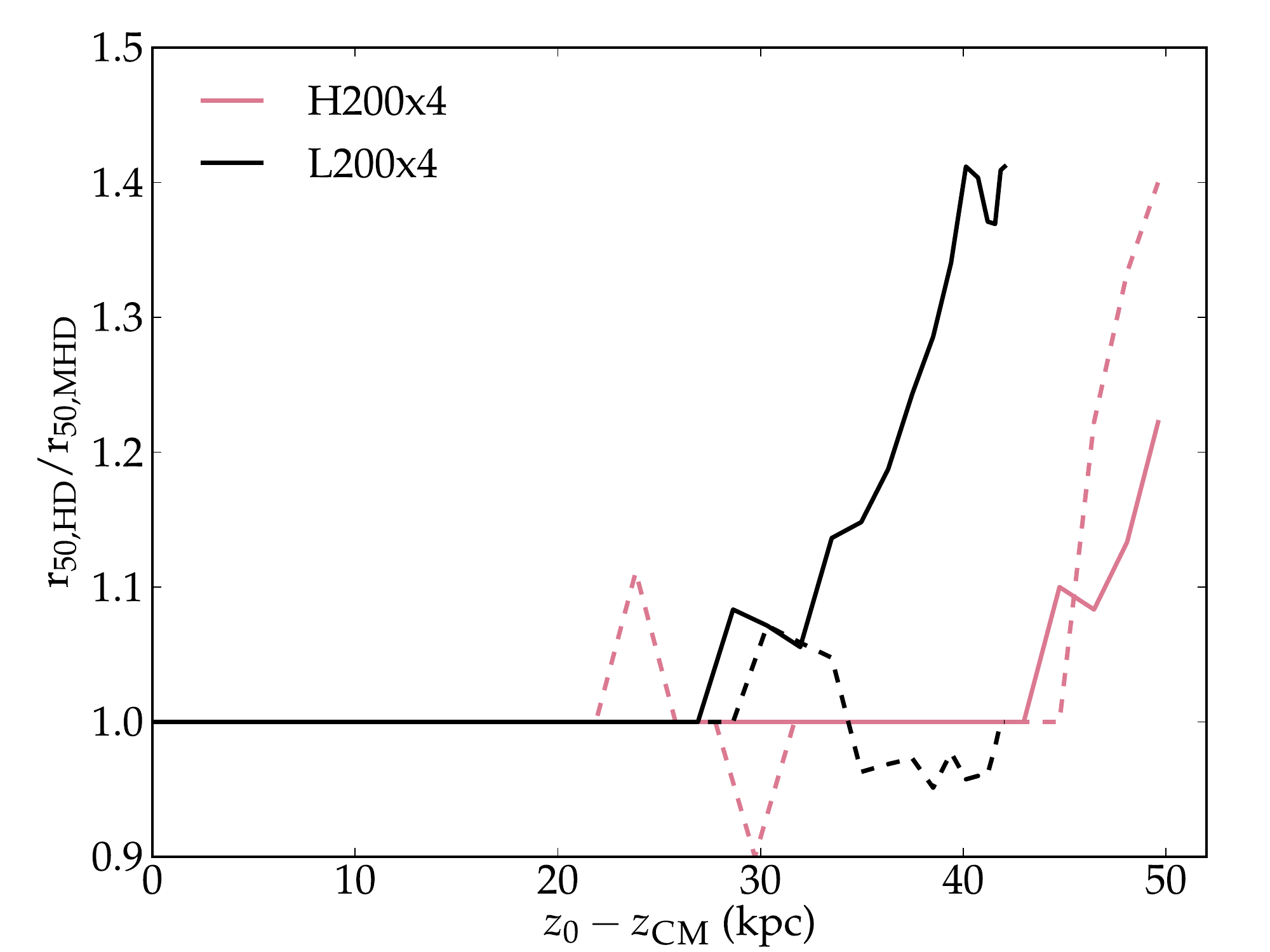}
\end{center}
\caption{Evolution of the cylindrical half mass radius along $x$ (solid lines) and $z$ (dashed lines) of clouds moving through a non-magnetized medium, relative to their magnetized counterparts.}
\label{fig:r50_hdmhd}
\end{figure}

The question of whether a gas cloud subjected to (magneto)hydrodynamic interactions has `survived' from a given initial state has no well defined answer, as it is not trivial to arrive at a robust definition of `survival'. A common approach is to quantify the evolution of the cloud's mixing with the ambient medium during its evolution. The intuition behind this is that the more a cloud mixes with the ambient medium, the more severe the ablation it has experienced. A cloud's mixing can be measured in different ways, for example, in terms of its density dispersion relative to its average density over a given volume volume \citep[e.g.][]{mccourt15}. Here, we introduce a different measure of a cloud's ablation, namely, its {\em half-mass radius}, i.e. the radius that encloses half of the mass of the initial cloud material at any given time. The idea is that the larger this radius, the more dispersed, and therefore the more `destroyed', a cloud. We define the half-mass radius as the radius of an infinitely long cylinder with symmetry axis along one of the coordinates centered on the cloud center of mass which contains half of the cloud's initial mass; we denote it by $r_{50,a}$ ($a \in  \{ x, \, y, \, z \}$). Note that measuring $r_{50,a}$ along different orthogonal axes is useful to separately assess the tail elongation and the transverse expansion. We define a second quantity, $r_{50,\text{sph}}$, as the radius of a sphere centered on the cloud center of mass encompassing half of the cloud's initial mass. Note that $r_{50,a}$ provides a more direct link to what can actually be measured. Indeed, radio observations can only provide spatial information on the plane of the sky, and are limited to kinematic information along the line-of-sight. 

We use $r_{50,\text{sph}}$ to estimate the evolution of the cloud-halo density contrast $\chi$ relative to the initial density contrast (see Figure \ref{fig:denscontrast}). Low density clouds experience an initial shock due to the supersonic wind. This compression in turn leads to a slight increase in the cloud's density relative to the background. Such an effect is also present in high density clouds, but it is weaker due to the lower Mach number of the shock. In all cases the density contrast declines monotonically throughout the majority of the cloud's journey toward the disk as the cloud is ablated. At $t \approx \tau_{cc}$ -- where the `cloud crushing time', $\tau_{cc} = 2r_c\sqrt{\chi}/v_{\text{wind}}$, is based on the initial density contrast and is roughly the time it takes for the shock to cross the cloud \citep{jones96}--, clouds experience significant disruption. This happens earlier for low density clouds, at $\sim 100 \Myr$. For the high density clouds $\tau_{cc}$ is higher by a factor of $\sqrt{5}$, so roughly 220 Myr. Cloud crushing times for the $v_{\text{wind}}=300\kms$ simulations can be found by dividing by 3/2. Note that the cloud crushing time is often defined with respect to the cloud radius rather than diameter in which case they will be a factor of 2 smaller.

The above holds true for clouds of similar density moving through a media with different densities. Indeed, making use of our newly defined metric, $r_{50,a}$, we find that clouds travelling through a less dense medium (L200x12) experience less ablation with respect to clouds in a high density environment (L200x4; see Figure \ref{fig:r50}). This is consistent with intuition, thus validating the use of $r_{50,a}$ as a quantitative measure of the survivability of HVCs moving through a magnetized halo. When the density of the medium is high enough (i.e. $n_h \sim 10^{-3} \pcc$ at $z_0 - z_{\text{\sc cm}} \approx 30 \kpc$), the cloud experiences significant ablation, as indicated by the dramatic increase in its half mass radius. High density clouds show a different evolution altogether. Indeed, regardless of the increasing density along their orbit, high density HVCs remain remarkably compact all along their journey. This is consistent with the fact that the cloud crushing time scale is longer for high density clouds compared to lower density clouds because of their greater density contrasts. There is no significant difference in the half-mass radii along transverse axes, as the half mass radii are dominated by the elongation along the tail in these cases. But there is a difference in the actual elongation along $x$ and $y$ at large $z_{\text{\sc cm}}-z_0$ with the dispersion of cloud material along $x$ being about 20 per cent higher than along $y$. This is a well known effect that occurs because the magnetic field inhibits RT instabilities along $y$ and $z$ and so the cloud expands more in the $x$ direction \citep{gregori99}. Our results suggests that for high density clouds, despite their elongated head-tail morphology \citep[][]{putman11}, much of the cloud's mass remains in the core. This, in turn, suggests that it is in fact not at all a poor approximation to estimate a cloud's (distance dependent) mass from its angular size alone.

As the reader may recall, we mentioned in the introduction that previous studies are somewhat inconclusive with respect to the question of whether magnetic fields prevent or enhance the destruction of HVCs. The most straightforward way we can address this is to compare simulations that include magnetic fields to  simulations without magnetic fields, but which are otherwise identical in their initial conditions. Such a comparison can be realized in a quantitative, objective way e.g. by calculating the evolution of $r_{50,a}$ in the purely HD case relative to the corresponding MHD case. We perform such a comparison for two runs with different initial cloud densities, $n_c=0.1 \pcc$ and $n_c=0.5 \pcc$, but fixed $x_0=4$ kpc and $v=200$ km s$^{-1}$. The result of this exercise is shown in Figure \ref{fig:r50_hdmhd}.

\begin{figure*}[h]
\begin{center}
\includegraphics[width=0.99\textwidth]{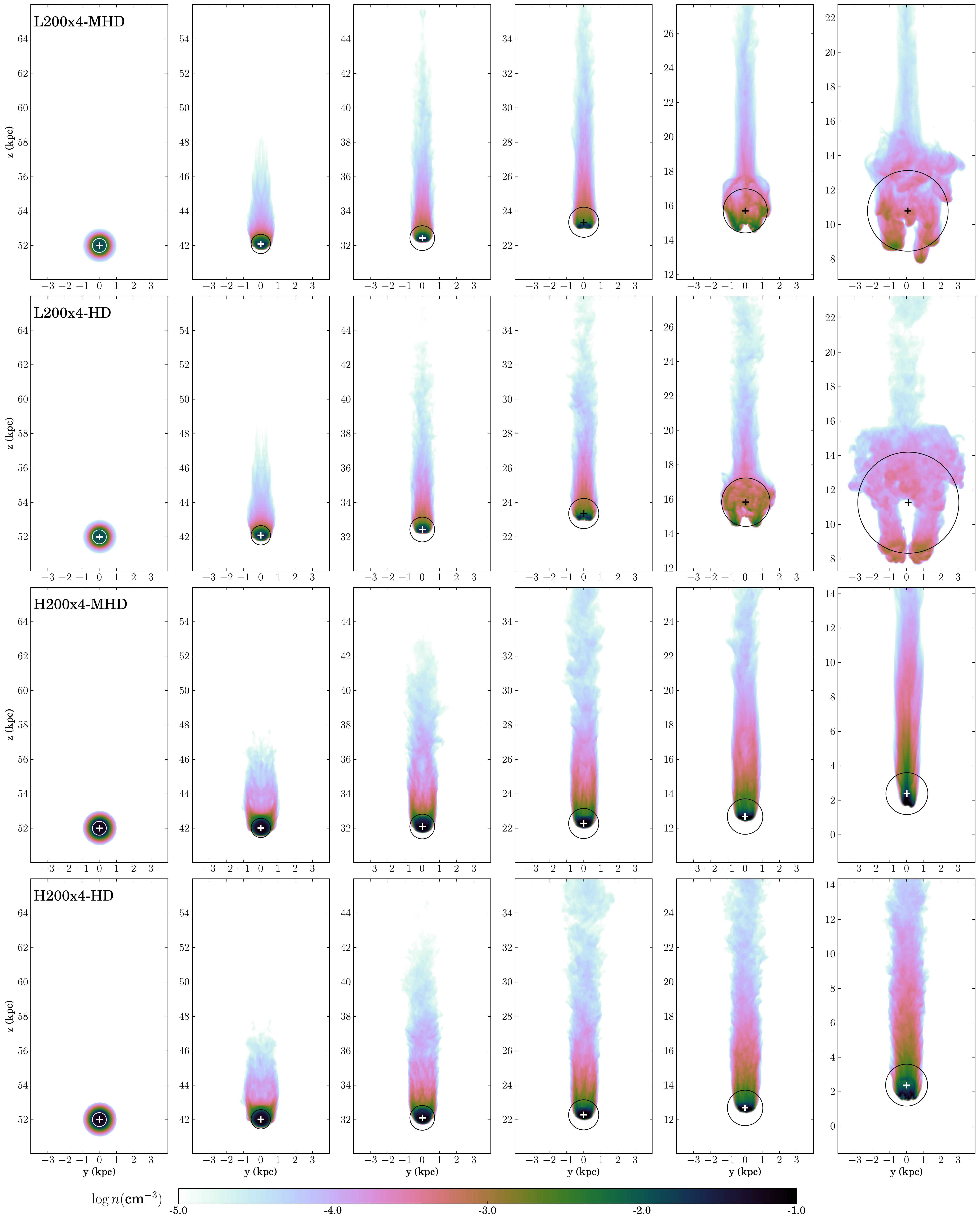}
\end{center}
\caption{Evolution of the cloud density projected along the $x$ axis. Note that the $z$-range varies across panels, from highest altitude on the left to lower altitude on the right, such that the cloud's center of mass stays close to the bottom. Here the entire density transition region $r<2r_c$ is included to emphasize the tail of material stripped from there. Each column corresponds to a different snapshot at $t = 0 \Myr$, 50 \Myr,  100 \Myr,  150 \Myr, 200 \Myr, and 250 \Myr, from left to right, respectively. Each row corresponds to a different simulation (L200x4, L200x4-HD, H200x4, H200x4-HD) from top to bottom. The cross and the circle in each panel indicates, respectively, the center of mass of the cloud and its spherical half mass radius. The color coding indicates the value of the density on a logarithmic scale.}
\label{fig:clouddens}
\end{figure*}

The first thing that becomes apparent is that clouds moving through a magnetized medium do tend to remain compact for a longer time, and more so if their density is high. In addition to breaking up earlier, low density clouds moving in a non-magnetic halo do so in a quite asymmetric fashion, being significantly more dispersed than their magnetized counterparts in the direction perpendicular to their orbit (and the ambient magnetic field). The behaviour is clearly shown in the density projections shown in Figure \ref{fig:clouddens}.

Low density clouds are essentially destroyed well before they get close to the Galactic plane, while high density clouds remain compact all the way to $z \approx 1.5 \kpc$ (at which point we stop the simulation). Low density clouds get destroyed despite the presence of an ambient magnetic field, although less severely than in its absence. High density clouds display a more compact morphology, and a stronger collimated tail, when moving through a magnetized medium. As noted before, their tail becomes coherent close to the Galactic plane as a result of the increased magnetic field in the halo there. In contrast, non-magnetized clouds display a diffuse, turbulent tail at all times. We did not show a comparison between HD and MHD simulations for the density contrast but they differ in the way expected from the current comparison. Throughout the majority of the simulations the density contrasts are similar but they become lower for the HD simulations at later stages with a roughly 50 percent difference relative to their MHD counterparts by the end.

A caveat to this analysis is that in our simulations the magnetic field is completely in the transverse direction which maximizes its amplification. However, while a magnetic field parallel to the cloud's direction of motion will lead to significantly less amplification, oblique fields angled at 45 degrees have been shown to lead to almost as much amplification as in the transverse case \citep{banda-barragan16}. We therefore conclude that, in fact, {\em magnetic fields delay the break-up of a cloud as it travels through the halo}, but perhaps not enough to guarantee that it reaches the disk.

\section{Discussion}

The qualitative evolution of the density and magnetic field of the cloud interacting with our time varying wind as seen in Figures \ref{fig:clouddens} and \ref{fig:2Dbmag} is broadly the same as in previous studies of clouds interacting with a constant wind with a uniform magnetic field  \citep[e.g.][]{gregori99,dursi08,banda-barragan16,goldsmith16}.

The distance from the Galactic plane at which specific magnetic field strengths are reached in the cloud is essentially independent of the cloud's initial velocity and density for distances reached by all clouds. However, denser clouds are able to travel further before being stopped by drag and thereby reach higher magnetic field strengths closer to the disk. Of course, with sufficiently high velocity a low density cloud would be able to travel as far but the required velocity to reach similar momentum is not realistic \citep{wakker91}. Additionally, low density clouds might not survive travelling that far (see Section \ref{sect:survival}). \emph{This indicates that the halo magnetic field is the dominant effect in the evolution of the cloud's magnetic field rather than the properties of the cloud.} This increases the usefulness of our model as it should be applicable to a wide range of HVCs including ones where little is known of their physical parameters. It also means that constraints on magnetic fields associated with HVCs might provide useful constraints on the halo magnetic field.

\subsection{Comparison with observations}

Magnetic field constraints are currently only available for two HVCs. However, with future surveys such as POSSUM \citep{gaensler10} we expect that magnetic field strengths will become available for a much larger amount. There are two primary methods for deriving constraints on magnetic fields from observations: rotation measures and Zeeman splitting.

The rotation measure is the measure of the change of polarization angle of emission observed at wavelength $\lambda$. It is proportional to the line-of-sigh integral of the product of the magnetic field parallel to the line of sight and the free electron density. Lower limits on the magnetic field strength along the line of sight have been derived from rotation measures in two HVCs: the Smith Cloud \citep{hill13} and an HVC in the Leading Arm \citep[HVC 287.5+22.5+240,][]{mcclure-griffiths10}. These studies assume a distance to the HVC, however this is only used to estimate the size of the cloud and rough limits can still be derived without assuming a distance. For the Smith Cloud, $B_{\vert\vert} \gtrsim 8$ $\mu$G is obtained. The location of the Smith Cloud is known to quite good accuracy, $z=-2.9 \pm 0.3$ kpc and $R=7.6 \pm 0.9$ kpc \citep{lockman08}, which is in agreement with the conclusion from our model that it must be within $\sim 10 \kpc$ of the Galactic disk. For HVC 287.5+22.5+240 $B_{\vert\vert} \gtrsim 6$ $\mu$G is obtained. Neither the distance to this HVC nor the distance to the Leading Arm II complex that it is a part of, is known. Nonetheless, it is assumed to be closer than the Large Magellanic Cloud, which has a well constrained Galactocentric distance of $d \approx 50$ kpc, with a vertical component $z=28$ kpc \citep{pietrzynski13}. Thus, the upper limit on its associated field is in agreement with our results.

Zeeman splitting causes a spectral line to split into multiple lines in the presence of a magnetic field. This effect can be used to estimate the magnetic field strength along the line of sight but it requires higher signal-to-noise ratios than the rotation measure method. As a result, no robust constraints on HVC magnetic fields have yet been derived using this method. An early promising candidate in the literature was HVC 132+23-212 \citep{kazes91} but this was later shown to be the result of an instrumental artifact \citep{verschuur95}. In short, \emph{of the two HVCs with available magnetic field constraints one is within a few kpc of the Galactic plane in agreement with our model for its relatively high field strength of at least several $\mu$G. The other one has no known distance but must be at $z<28$ kpc and is thus consistent with our model which predicts that this HVC should be close to the disk.}

The underlying assumption for this work is that the measured magnetic field corresponds to the ambient field that has been swept up and amplified at the cloud's leading edge. But this interpretation may certainly not apply in general. Indeed, in some cases such as the magnetic field recently measured in the Magellanic Bridge \citep[$B_{\vert\vert}=0.3\pm 0.3$ $\mu$G;][]{kaczmarek17}, it is more likely to represent magnetic field that has been pulled from the Magellanic clouds. In cases where distances are already available through a more reliable method, our method can be used to examine the origin of the cloud's observationally derived magnetic field by assessing whether the field is consistent with being swept up halo field or if it must have another source.

Figure \ref{fig:headvstail} is especially relevant for observations as it provides a prediction of a \emph{relative} measure of the variation of the magnetic field across the cloud and so is independent of instrumental artifacts and of a detailed knowledge of the ambient field.

\subsection{Convergence}
\label{sect:convergence}

\begin{figure}[h]
\begin{center}
\includegraphics[width=0.49\textwidth]{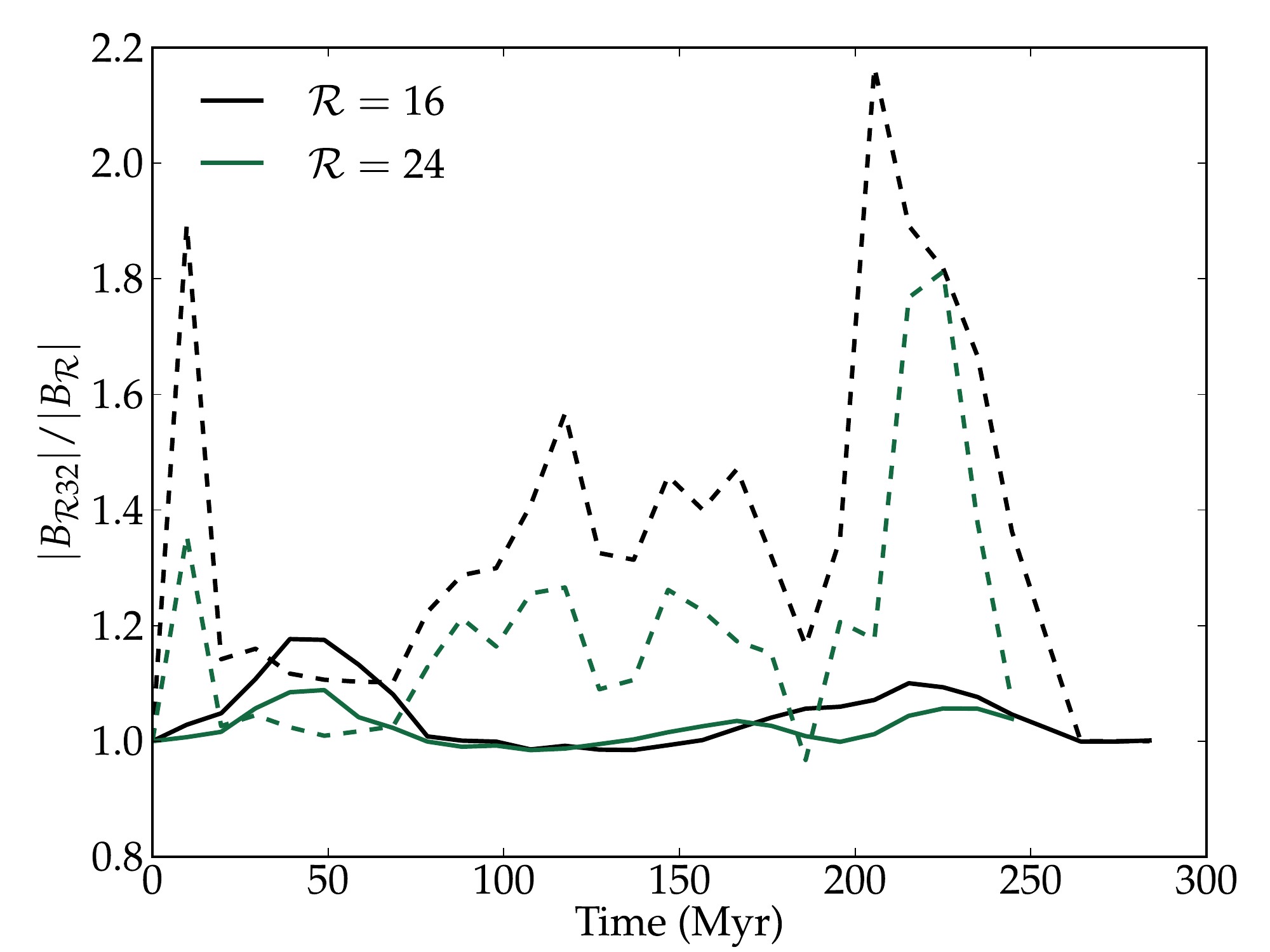}
\end{center}
\caption{Evolution of \Bnine\ (solid) and \Bmax\ (dashed) at a resolution of 32 cells per cloud radius relative to their evolution in two lower resolution runs.}
\label{fig:convergence}
\end{figure}

In order to assess whether our standard resolution of 16 cells per cloud radius, $\mathcal{R} \equiv r_c/\Delta x=16$, is sufficient to resolve the amplification of the magnetic field around the cloud, we perform a crude convergence test. More specifically, we compare the evolution of \Bnine\  and \Bmax\ in runs with different spatial resolution. To this end, we re-run simulation L200x4 with 1.5 and 2 times the default resolution, i.e., $\mathcal{R}=24$ and $\mathcal{R}=32$, respectively. A quantitative comparison is achieved by computing the ratio of a given quantity at our highest resolution $\mathcal{R} = 32$,  say \Bnine, (denoted by $\vert B\vert_{90,\mathcal{R}32}$), with the corresponding quantity at a different resolution, $\vert B\vert_{90,\mathcal{R}}$,. We show such a comparison in Figure \ref{fig:convergence}. Clearly, $\vert B\vert_{90,\mathcal{R}32}$ is only slightly higher than $\vert B\vert_{90,\mathcal{R}24}$, and the latter only slightly higher than $\vert B\vert_{90,\mathcal{R}16}$ at all times, except perhaps around $t\approx 150$ Myr. This indicates that \Bnine\ is converging although it has not fully converged yet at $\mathcal{R} = 32$. \Bmax\ does not show convergence as good as \Bnine\, although this is not surprising as it a one-pixel statistics, thus subject to large fluctuations, in contrast to  \Bnine\ which is robust. Nonetheless, $\vert B\vert_{\text{max},\mathcal{R}16}$ and $\vert B\vert_{\text{max},\mathcal{R}24}$ remains within about a factor of two at all times. We conclude that a standard resolution of $\mathcal{R}=16$ is sufficient to carry on experiments like ours.

It is worth mentioning that \cite{dursi08} found that, in order to resolve the magnetic field at the cloud's leading edge in their simulations, at least $\mathcal{R}=32$ was necessary. Likewise, \cite{goldsmith16} found that the cloud mass and velocity was reasonably converged at $\mathcal{R}=32$ as well. But care must be taken when comparing this type of simulations at different resolutions. Resolutions are usually stated in terms of the cloud radius, however definitions of this differ as cloud density profiles vary between studies. Typically the density profile is smooth. It might consist of an inner region of essentially constant density surrounded by a transition layer, as in our simulations, or it might decline immediately outside $r=0$. For instance, \cite{dursi08} and \cite{banda-barragan16} both use profiles with wide transition layers and include these in the cloud radius (in fact our profile is similar to the one used in \cite{banda-barragan16} except that they set the steepness to $s=10$). If the radius is taken to follow our definition $n(r_c) \approx n(0)/2$ then the radius and thus resolution in these studies should be halved. \cite{goldsmith16} also included the entire transition region in their definition of the cloud radius, however this transition region was narrower than in the previously mentioned studies. When the differences in the density profiles and definitions of cloud radius are taken into account our result that the magnetic field amplification is reasonably converged at a resolution of $\mathcal{R}=16$ is consistent with previous results.

\subsection{Limitations} \label{sect:limitations}

There is a degeneracy in the magnetic field strength between the radius in the Galactic plane, $R$ (denoted by $x$ in our simulations), and the distance from the Galactic plane, $z$. However, comparing the amplification of the halo field in the cloud for our $x_0=4$ kpc and $x_0=12$ kpc simulations (see Figure \ref{fig:bampvshalo}) shows that it is roughly equal. If we assume that this holds for all $R$ then we can generate a full three dimensional model of cloud magnetic field strengths in the galaxy and upper limits on $R$ and $z$ can be estimated from the point where the line of sight crosses the contour of the upper limit of the magnetic field strength. At this stage, however, we are only concerned with constraining distances to within a factor of several mainly to establish whether a given HVC is relatively `near' or `far'.

The amplification of the magnetic field in front of the cloud depends on the angle of the field lines with respect to the cloud's trajectory. The reference orbit we have adopted in all our simulations -- i.e. perpendicular to the Galactic plane -- corresponds to an extreme case where the halo magnetic field is everywhere (roughly) perpendicular to the cloud's motion, which in turns yields the maximum possible amplification. However, an oblique field at an angle of 45 degrees in fact leads  to an amplification similar to our maximal case, albeit for clouds moving through a {\em uniform} medium \citep{banda-barragan16}. 

As mentioned in Section \ref{sect:ICs}, we ignore the radial variation of the magnetic field and density of the halo. The density and strength and direction of the magnetic field changes significantly across the 8 kpc$^2$ $xy$ plane of our simulations. However, only the variations across the cloud is important for our purposes. These are relatively small over the 0.5 kpc initial radius of the cloud. However, the clouds do increase in size during the simulations (see Section \ref{sect:survival}) so to check that our approximation is valid throughout the simulations we ran L200x4-rv, H200x4-rv and L200x12-rv in which the radial variation is included. We found no significant differences between these three simulations and their radially invariant counterparts.

Our simulations include many simplifications. A significant omission is not including the gravity of the Galaxy. The simplest way to do this would be to assume some dark matter density profile and then calculate the gravitational acceleration of the cloud's center of mass from the static potential of the sum of the dark matter and halo gas density profiles (more components could be included for increased accuracy, e.g. \citealt{mcmillan17}). This would however complicate the analysis considerably. The acceleration would pull the cloud horizontally in the $x$ direction as well as vertically which would lead to the clouds having different paths depending on $x_0$. It would also mean that our approximation that the magnetic field strength and density only depends on $z$ would no longer be appropriate. With a gravitational potential included and only taking the velocity into account, the low density clouds would presumably be able to reach the disk rather than being stopped by drag at $z \approx 10$ kpc. However, based on our results they would probably be destroyed before reaching the disk.

We use the ideal MHD approximation to evolve the magnetic field in the simulations. In doing so we assume that the gas is sufficiently ionized to be described as a single fluid with negligible ambipolar diffusion and diamagnetic current terms \citep[see][]{pandey08}. The halo gas in the simulations has temperatures $T \gtrsim 10^4$ K at all times and can therefore be assumed to be highly ionized. The halo temperature would drop below $10^4$ K for simulation L200x4 at the smallest allowed distance from the plane of $z=1.5$ kpc, however the cloud is stopped by drag before reaching parts of the halo at these temperatures. The mass of ionized gas in HVCs is comparable to the HI mass \citep{putman12} and surrounding ionized gas is observed in e.g. the Smith Cloud \citep{hill09} and Complex C \citep{fox04,collins07}. Thus we would expect the ionization fraction of an HVC to be high in the outskirts and low in the core. In our simulations we are concerned with the magnetic field amplification which occurs in front of the cloud and in the tail behind the cloud and the magnetic field remains weak compared to the halo field in the inner parts of the cloud (see Figure \ref{fig:2Dbmag}). Thus our assumption of high ionization fraction is reasonable in the regions of interest in our simulations.

Hall drift is an effect that for a fully ionized gas is caused by the difference in inertia between  ions and electrons. The length scale at which Hall drift becomes significant is $L_H=\frac{v_A}{\omega_H}$ where $v_A=B/\sqrt{4\pi\rho}$ is the Alfv\'{e}n velocity and $\omega_H$ is the Hall frequency \citep{pandey08}. For fully ionized low-metallicity gas $L_H\approx 1.8 \times 10^{-5}$ cm$^{-1/2}$ g$^{1/2} \times \rho^{-1/2}$ ($B$ cancels out as $\omega_H \propto B$). For the densities present in our simulations this is much less than the smallest length scale $\Delta x\sim 10$ pc that is resolved in our highest resolution simulation. The further assumption of ideal MHD is that the resistivity is negligible. The validity of this assumption can be assessed through the magnetic Reynolds number $R_m=vL/\eta$ where $v$ and $L$ are typical velocity and length scales and $\eta$ is the resistivity which approximately depends on the temperature through $\eta \sim T^{-3/2}$ \citep{spitzer53}. Even at the smallest length scale we can resolve in the highest resolution simulation and the lowest temperatures present, the initial temperature in the cloud, $R_m \gg 1$ for the velocities considered in our simulations (including when the velocity is chosen to be the Alfv\'{e}n velocity and $R_m$ becomes the Lundquist number) indicating that resistivity is relatively unimportant. Our simulations do however have numerical resistivity as do all grid-based MHD simulations analogous to numerical viscosity \citep[see e.g.][]{fromang07}. Non-ideal MHD effects can still be important for the field topology through magnetic reconnection, however in this paper we only consider the amplitude of the magnetic amplification around the cloud and the overall effect of the magnetic field on cloud survival.

Finally, although we are aware that radiative cooling may be important for simulations like ours \citep[see, e.g.,][]{mellema02,cooper09,scannapieco15}, we have chosen to ignore this process for now. Including an advanced treatment of cooling -- using e.g. {\sc mappings} \citep{sutherland93} or {\sc cloudy} \citep{ferland13} -- in three-dimensional simulations with reasonable resolution is computationally costly, but possible. Doing so renders simulations like ours more realistic, but it also introduces a high level of complexity which may obscure other effects. Thus, for the sake of clarity, we have run our simulations adiabatically, deferring the more challenging task to include cooling for future work. In addition to cooling, other physical effects that may be relevant when modelling cloud-wind interactions include: thermal conduction \citep{armillotta16,bruggen16}; turbulence \citep[e.g][]{pittard16}; and fractal density structure \citep[e.g.][]{bland-hawthorn07,schneider17}. Taking all these effects into account in MHD simulations is computationally expensive, often limiting the geometry to two dimensional simulations, which may be even less realistic that full 3D simulations ignoring these processes.

\section{Conclusions}

If the magnetic fields on the order of $\mu$G associated with HVCs are in fact {\em not} intrinsic to the cloud but rather corresponds to the ambient field that has been `swept up' by the cloud along its orbit, then our study suggests that such HVCs are relatively close ($z \lesssim 10 \kpc$) to the disk of the Galaxy. This is in agreement with the two HVCs that have observational constraints on their magnetic field strengths and distances. In other words, we suggest that measurements of magnetic fields around gas clouds in the vicinity of the Galaxy may be useful to put an upper limit on their distance.

In addition, our results suggest that magnetic fields could in fact delay the destruction of gas clouds by hydrodynamic instabilities. Although the effect of the halo magnetic field is fairly limited mainly because hydrodynamic effects dominate by far along most of a cloud's journey, close to the Galactic plane, magnetic fields become dynamically important, and clouds tend to remain compact for a longer time. We defer a systematic study of the impact of magnetic fields on cloud survival to a forthcoming paper.

\section*{Acknowledgements}
AG and TTG acknowledge financial support from the Australian Research Council (ARC) through an Australian Laureate Fellowship awarded to JBH. NMG acknowledges the support of the ARC through Future Fellowship FT150100024. We thank Wladimir Banda-Barrag\'an for assistance on numerical methods.

This research was undertaken with the assistance of resources and services from the Sydney Informatics Hub and the University of Sydney's high performance computing cluster Artemis; the National Computational Infrastructure (NCI), through the Astronomy Supercomputer Time Allocation Committee scheme managed by Astronomy Australia Limited and supported by the Australian Government; and the gSTAR national facility at Swinburne University of Technology. gSTAR is funded by Swinburne and the Australian Government's Education Investment Fund.

\software{PLUTO \citep[version 4.1 of the code last described by][]{mignone07,mignone12}; VisIt \citep{HPV:VisIt}}.

\bibliography{AG_bfield_distance_constraint}

\end{document}